\documentclass[letterpaper,9pt]{article}
\pdfoutput=1

\usepackage{geometry}
\usepackage{authblk}
\usepackage{booktabs}
\usepackage{multirow}
\usepackage{graphicx}
\usepackage{amsmath}
\usepackage{amssymb}
\usepackage{float}

\title{A Compressed Sampling and Dictionary Learning Framework for WDM-Based Distributed Fiber Sensing}

\author[1]{Christian Weiss}
\author[1]{Abdelhak M. Zoubir}

\affil[1]{Signal Processing Group, Institute of Communications and 
       Graduate School of Computational Engineering, Technische Universit\"at Darmstadt, Darmstadt, Germany}
\date{\small\today}

\newcommand\blfootnote[1]{%
  \begingroup
  \renewcommand\thefootnote{}\footnote{#1}%
  \addtocounter{footnote}{-1}%
  \endgroup
}

\begin{document}

\maketitle

\begin{abstract}
\noindent
We propose a compressed sampling and dictionary learning framework for \mbox{fiber-optic} sensing using wavelength-tunable lasers.                                                    								  
A redundant dictionary is generated from a model for the reflected sensor signal. Imperfect prior knowledge is considered in terms of uncertain local and global parameters.                
To estimate a sparse representation and the dictionary parameters, we present an alternating minimization algorithm that  is equipped with a pre-processing routine to handle           			  
dictionary coherence. The support of the obtained sparse signal indicates the reflection delays, which can be used to measure impairments along the sensing fiber.  						   			  
The performance is evaluated by simulations and experimental data for a fiber sensor system with common core architecture.      			
\blfootnote{Accepted for publication in Journal of the Optical Society of America A [ \copyright\ 2017 Optical Society of America.\,] One print or electronic copy may be made for personal use only. Systematic reproduction and distribution, duplication of any material in this paper for a fee or for commercial purposes, or modifications of the content of this paper are prohibited. }   
\end{abstract}

\section{Introduction}
Fiber-optic sensors have paved the way to a wide range of sensing applications based on different sensing \mbox{technologies 
\cite{Udd1996,Kersey1997,Culshaw2008}.}  %
Fiber Bragg grating (FBG) sensors are often employed in smart structures for high-resolution quasi-distributed temperature or strain monitoring \cite{Culshaw2004,Hotate2008,Murayama2013}.
In systems based on wavelength-division multiplexing (WDM), fiber interrogation is performed using wavelength-tunable lasers that feature high local signal-to-noise ratios (SNRs).
For quasi-distributed sensing, a wide tuning range is necessary, besides high sweep rates, to monitor time-varying perturbations such as vibrations \cite{Nakazaki2009,Yamashita2009}. %
In order to infer the quantity or nature of impairments at the FBGs, precise estimation of the reflection delays is important \cite{Culshaw2004,Nakazaki2009,Yamashita2009}.\\
Constant monitoring at high sweep rates results in large amounts of data to be stored/processed and requires costly high-speed analog-to-digital converters (ADCs). 
Compressed Sampling (CS) is an emerging technology that addresses these problems \cite{Baraniuk2007,Candes2008,Eldar2012}. 
The measurements in CS are taken in form of low-dimensional random projections. 
In order to recover the original signal or to extract the desired information, an ill-posed inverse problem has to be solved.  
By relying on the sparsity of the signal with respect to some known dictionary, efficient recovery algorithms have been developed, which is the basis for the success of CS \cite{Eldar2012,Foucart2013}.
Therefore, CS has become a favorable tool in various sensing \mbox{applications \cite{Malioutov2005,Yu2009,Herman2009,Lustig2008,Patel2010,Foucart2013,Gehm2015}.}\\  %
A sparsity-promoting dictionary is the key component in any sparse signal processing method such as CS. 
Popular dictionaries are based on a classical Fourier or wavelet analysis \cite{Candes2008}. 
Nevertheless, redundant dictionaries gain increasing importance as they are more likely to yield a sparse representation \cite{Schnass2008}. 
In dictionary learning (DL), the acquired data is used to create a dictionary. 
Parametric DL (PDL) approaches consider a dictionary that is created from a mathematical model. They use the acquired data to adapt the dictionary parameters. 
Alternating \mbox{minimization (AM)} is a prominent heuristic for iteratively estimating a dictionary along with a sparse representation \cite{Arora2014,Agarwal2014}. 
A general requirement for the success of CS is that the columns of the sensing dictionary exhibit low \mbox{coherence \cite{Elad2007,Schnass2008},}  %
where the term \emph{coherence} is used to describe the level of similarity between dictionary columns (called \emph{atoms}).
For redundant dictionaries, pre-processing helps to alleviate this \mbox{problem \cite{Candes2011}}.\\  
In this work, we propose a versatile unified CS and DL (CS-DL) framework for an automated estimation of the reflection delays and dictionary parameters 
in WDM-based fiber sensing. 
In the CS-DL framework, we address several aspects with respect to the targeted application. 
First, we apply CS to reduce the sampling rate and the number of samples to be stored and processed. 
Second, we create an application-specific sparsity-promoting dictionary based on a model for the received sensor signal, which is obtained by compiling established mathematical models for the individual system components.
Sparsity, in this context, corresponds to the small number of FBGs in the sensing fiber with respect to the fiber length. 
The total received signal contains one reflection from each FBG. Therefore, this signal can be sparsely represented in a dictionary of incrementally translated reflections. 
Fourth, we estimate unknown model parameters using PDL. 
Fifth, we use an orthogonal matching pursuit (OMP) algorithm to estimate a sparse representation of the signal, which yields estimates of the desired reflection delays.
To obtain a reliable sparse representation, we address the problem of dictionary coherence using inter-atom interference (IAI) \mbox{mitigation.}

\subsection{Contributions}
\begin{itemize}
 \item[(I)]   
			We provide a versatile CS-DL framework for WDM-based quasi-distributed fiber-optic sensing that pairs the advantages of CS and PDL. 
			The purpose of CS is to abate the sampling rate and to reduce the number of samples to be stored/processed, 
			while the purpose of PDL is to handle incomplete prior knowledge of the signal by estimating unknown model parameters.   %
 \item[(II)] We  identify a generic parametric dictionary with local and global parameters to estimate the reflection delays directly in the sparse domain. 
			 As a figure of merit, we introduce a novel, auxiliary coherence measure, called the \emph{coherence distance}. Its purpose is to evaluate the level of difficulty of the sparse estimation problem for 
			 different parameter settings at high sparsity levels. This measure can be widely used for general shift-invariant dictionaries. 
			 To emphasize this, we also establish a relation to an existing popular measure, the Babel \mbox{function \cite{Tropp2004}.}
 \item[(III)] 
 			 We introduce an algorithm for sparse estimation and DL, called PDL-OIAI. %
			 Its purpose is to estimate the reflection delays along with unknown model parameters based on a highly redundant dictionary. 
			 PDL-OIAI adopts the IAI mitigation method in \cite{Yang2010} to handle dictionary coherence. 
			 We show that IAI mitigation allows us to use a sub-optimal greedy algorithm for sparse estimation. %
			 We also analyze the computational complexity of PDL-OIAI. %
 \item[(IV)] We show how to tailor the general CS-DL framework to match a specific system setup for WDM-based fiber sensing. 
 			To this end, we use existing mathematical models for the individual sensor components to compile an overall model for the observed signal.
 			We use this model to create an \mbox{application-specific} dictionary. 
 \item[(V)]  We present simulations to assess the performance of PDL-OIAI %
 		     for various scenarios of different dictionary parameter values, CS sampling matrices, and SNRs. 
 		     Finally, these results are supported and verified using experimental data from a real fiber sensor system in \cite{Nakazaki2009,Yamashita2009}. %
\end{itemize}

\subsection{Related work}
To date, only a few sources report the application of CS to WDM-based distributed fiber sensing. This work refines and extends some prior investigations in \cite{Weiss2013,Weiss2015}. 
The authors in \cite{Lin1998,Mishali2009} describe the CS acquisition process in terms of a periodic non-uniform sampling paradigm.
Based on this work, we describe non-uniform sampling by a sparse CS sampling matrix, using the \emph{Database-Friendly} (DF) distribution \mbox{in \cite{Achlioptas2003}.}\\ 
The concept of AM-based estimation (i.e. alternating minimization) was introduced \mbox{in \cite{Olshausen1996}} and, later on, generalized in \cite{Lewicki2000}. 
This has motivated the development of various other AM-based DL approaches. 
Please refer to \cite{Tosic2011} for a review. In many existing approaches, the entire dictionary atoms are learned.
%
Our work, in contrast, aims at estimating the parameters of a model-based dictionary, which is created by considering the underlying physical 
processes and the quantities to be estimated. %
We closely follow the experimental setup and considerations in \cite{Nakazaki2009,Yamashita2009}.
Model-based dictionaries have been previously investigated for different applications in \cite{Liang2013,Yao2015}.
Another design method \mbox{in \cite{Ataee2010}} creates the dictionary based on minimum-coherence constraints.\\   %
Redundant dictionaries have been analyzed in the context of \mbox{CS \cite{Candes2011}}.
The general problem of dictionary coherence in sparse estimation is investigated in \cite{Elad2007,Schnass2008}, where optimal sensing dictionaries are derived with respect 
to coherence measures such as the \emph{mutual coherence} and the \emph{Babel function}. %
However, it is found that the performance gain in these methods is limited because no information of the acquired data is \mbox{used \cite{Elad2007,Yang2010}.} 
In PDL-OIAI, we exploit such information by applying an adaptive method for IAI mitigation, proposed \mbox{in \cite{Yang2010}.}\\
For sparse estimation and PDL, different approaches \mbox{exist \cite{Austin2010,Austin2013,Fyhn2015,Leigsnering2016,Raja2016}.}
The method in \cite{Austin2010} aims at high-resolution parameter estimation with joint model order estimation using common information criteria. 
Based on this approach, the method in \cite{Austin2013} can additionally adjust the dictionary parameters according to the measured data. 
The effect of parameter quantization due to a finite number of dictionary atoms is considered in \cite{Fyhn2015}, where a polar interpolation between adjacent dictionary atoms is applied. 
The method in \cite{Austin2013} is similar to PDL-OIAI in that an AM scheme is used to successively estimate, first, the sparse
coefficients and, subsequently, the dictionary parameters by solving a least-squares optimization problem with coherence constraints.  %
In PDL-OIAI, however, we do not explicitly restrict the coherence but include an IAI mitigation step for dictionary pre-processing.
The PDL approach in \cite{Leigsnering2016} considers uncertain dictionary parameters in terms of unknown wall-locations in through-the-wall-radar imaging.  %
This technique is extended \mbox{in \cite{Raja2016},} where a network of radar systems is employed to estimate an indoor scattering scene using AM. 
In contrast to PDL-OIAI, no dictionary pre-processing is applied.

\subsection{Outline} 
Section \ref{sec:working_principle} introduces the working principle of WDM-based quasi-distributed fiber sensing.
Also, a generic parametric dictionary for sparse estimation is introduced. 
Section \ref{sec:CFS} describes the use of CS in our application along with recovery guarantees and coherence measures.  
In Section \ref{sec:paramDictLearn}, the CS-DL framework is introduced. Also, the PDL-OIAI algorithm is presented and its computational complexity is analyzed. 
Next, in Section \ref{sec:CS_DL_pratctice}, the CS-DL framework is truncated to suit a particular system setup by compiling a
parametric signal model and specifying a suitable dictionary. 
The performance of PDL-OIAI is evaluated in simulations and experimentally validated using real data. 
A discussion of the results and findings is provided in Section \ref{sec:discussion}.
Section \ref{sec:conclusion} concludes this work.

\section{System Setup and Sensing Principle}	%
\label{sec:working_principle}
\begin{figure*}[htbp]
  \centering
\includegraphics[width=1.0\linewidth]{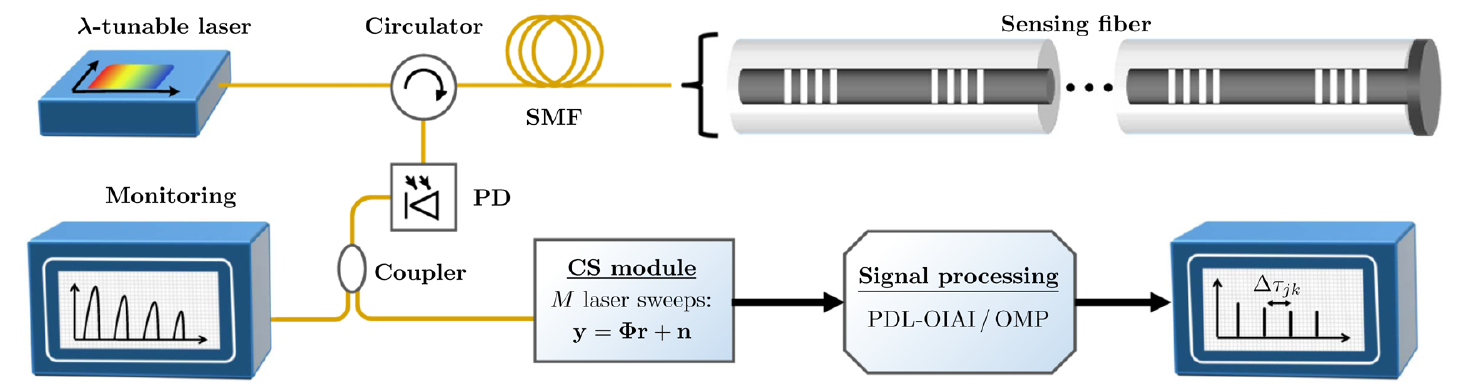}\vspace{-0.1cm}
   \caption{Basic system layout for WDM-based quasi-distributed fiber-optic sensing with CS-based acquisition, followed by sparse estimation and 
            DL using the PDL-OIAI or PDL-OMP algorithm.} \vspace{-0.2cm}
  \label{fig:cfsp}
\end{figure*}   
At first, we describe the essential components of the considered core architecture based on the fiber sensing system in \cite{Nakazaki2009,Yamashita2009}. %
We assume a modular architecture in which additional components may be added. %
Then, we review the sensing principle and specify the quantities to be retrieved from the CS samples.

\subsection{Basic setup}
Let us consider the system in Fig.\ \ref{fig:cfsp} which is based on the experimental setup in \cite{Nakazaki2009,Yamashita2009}.
The core system is extended by a CS acquisition module and a signal processing block.
The employed wavelength-tunable laser is assumed to operate in pulsed mode. 
The emitted signal travels through a single-mode fiber (SMF) into the sensing fiber. 
If the emitted wavelength falls into the reflection bandwidth of an FBG, it is reflected back to the receiver.
A circulator redirects the signal towards the photodetector (PD), where it is converted to the electrical domain.
For the purpose of visualization, a coupler may send one portion of the signal, 
e.g. $10\%$, to a monitoring unit and the rest to the CS acquisition module. Finally, the desired information is extracted from the acquired CS samples using PDL-OIAI (see Section \ref{sec:paramDictLearn}).  \vspace{-0.2cm}

\subsection{Impairment monitoring using FBGs}
FBGs reflect light within a narrow region around their Bragg wavelength, $\lambda_B = 2n_{\text{\scriptsize eff}}\Lambda_{\text{\tiny FBG}}$ \cite{Morey1990}, where $n_{\text{\scriptsize eff}}$ is the effective refractive index 
of the propagating mode and $\Lambda_{\text{\tiny FBG}}$ is the grating period (grating pitch). 
They are transparent to other wavelengths. 
The reflection spectra of the FBGs along the sensing fiber are assumed to have a well-separated spectral support and may exhibit a non-uniform shape. 
Any impairments at one FBG are assumed to be uniformly distributed along its spatial extent. 
For vibration monitoring, this is fulfilled if the acoustic wavelength of a vibration signal is much larger than the grating length \cite{Culshaw2008a}. 
Then, the grating period is uniformly stretched/compressed, 
resulting in a shift of $\lambda_B$ by $\delta\lambda_B$ which indicates the effective quantity of impairments. 
This shift can be caused by strain, $\epsilon_s$, or temperature change, $\Delta T$, yielding %
\begin{equation}
\label{eq:phys_rel_shift}
  \frac{\delta\lambda_B}{\lambda_B} = P_e\epsilon_s + [P_e(\alpha_s-\alpha_f) + \eta_t]\Delta T,
\end{equation}
where $P_e$ is the strain-optic coefficient, $\alpha_s$ and $\alpha_f$ are the thermal expansion coefficients of the bonding material and the fiber, respectively, and
$\eta_t$ is the thermo-optic  \mbox{coefficient \cite{Othonos1999,Kashyap2009,Yin2008}}. Changes in strain and temperature are \emph{a priori} indistinguishable, unless the strain component varies considerably faster than the temperature. 
However, if the temperature and strain coupling coefficients of two adjacent gratings are known, it is possible to recover both quantities without ambiguity \cite{Yin2008}.
\subsection{Quantifying impairments from time delays}%
\label{sec:quantifyingImpairments}
When the lasing wavelength is swept through the entire wavelength region, $R_{\text{sw}}$, of the gain medium, pulses at varying center wavelength 
are launched into the sensing fiber. Suppose that there are $K$ FBGs in the sensing fiber. The single pulses are reflected if their spectral support overlaps with the reflection region of FBG$_{k},\ k=1,\dots,K$.
In order to avoid ambiguities, we require that the spatial distance of the FBGs to the receiver increases with $k$, and that $\lambda_{B,k} \ll \lambda_{B,{k+1}}\ \forall\ k$.
Then, each reflection can be uniquely assigned to a single FBG.
Due to a limited bandwidth of the PD and the receiver circuitry, the individual laser pulses are usually not resolved and only the envelope modulating 
the pulses is observed. Let us use FBG$_{j},\ j\in\{1,\dots,K\},$ as a reference point, which is located in a controlled environment and never subject to perturbations. 
For a constant sweep rate, $S_{\text{r}}$, the reflected envelopes arrive with delays $\tau_{k},\, k=1,\dots,K$, where $\tau_{j}$ is the delay associated with the reference point.
Then, 
\begin{equation}
\label{eq:referenceDelay}
 \Delta\tau_{jk} = \left|(\tau_k\pm\delta\tau_k) - \tau_j\right|\ , %
\end{equation}
is the differential delay and $\delta\tau_k$ is the absolute shift induced by impairments acting at FBG$_k$. %
By precisely estimating $\delta\tau_k$ (or $\Delta\tau_{jk}$), we can infer $\delta\lambda_k\ \forall\ k\in\{1,\dots,K\}\setminus j$ by \cite{Nakazaki2009}
\begin{equation}
 \delta\lambda_{B,k} = \delta\tau_k \, S_{\text{r}} \  R_{\text{sw}} \ .
\end{equation}
The amount of strain or temperature change is then obtained \mbox{from (\ref{eq:phys_rel_shift})}.
When time-varying perturbations are monitored, $S_{\text{r}}$ has to be much faster than the perturbation frequency. 
However, we assume that the observed scene is stationary during the CS acquisition process. 

\subsection{Generic parametric dictionary for delay estimation}  
\label{sec:sparse_dict}
In order to identify an appropriate dictionary for estimating the reflection delays, we have to find an expression for the sensor measurements. 
Let us first define the observation time per laser sweep by $T_{\text{sw}}$. %
Next, we divide the observation time interval  into $K$ non-overlapping segments (according to the number of FBGs, $K$).  %
Each segment represents the considered delay range of a signal reflected from a certain FBG.
The $k$-th delay segment is discretized and represented by a finite set of delays points, $\mathcal{T}_k$, $k=1,\dots,K$. 
It defines all possible reflection delays that are associated with a signal reflected from  \mbox{FBG$_k$, $k=1,\dots,K$.} 
In order to create a dictionary, each discrete delay is uniquely assigned to one dictionary entry (atom). We consider a number of $N$ dictionary atoms.
Let $\Omega = \{1,\dots,N\}$ be the total set of dictionary indices, and let $\mathbf{A}_k$ denote the dictionary matrix associated with a reflection from the $k$-th FBG. 
Then, the index subset associated with $\mathbf{A}_k$ is  
\begin{equation}
    \Omega_k=\{n\in\Omega\,|\, N_{k-1} \leq n \leq N_{k}\},\quad k=1,\dots,K,
\end{equation}
where $N_{k+1} > N_k$. The index of the first and last dictionary atom is given by $N_0 = 1$ and $N_K = N$, respectively.
We accredit the same number of grid points (delays) to each segment and choose $N$ such that  \mbox{$|\Omega_k|=(N/K)\in\mathbb{N}_+\ \forall\ k=1,\dots,K$.}\\
To specify the dictionary atoms, let $r_k(t),\ k=1,\dots,K$, denote the (continuous) reflection signal from FBG$_k$, centered around $t=0$.
At this point, we can think of $r_k(t)$ as a single pulse-like signal, e.g. a Gaussian pulse. It is formed by the directly detected optical power that is reflected from FBG$_k$ during the fiber interrogation.  
The overall received sensor signal is composed of $K$ reflections from $K$ FBGs, i.e.
\begin{equation}
\label{eq:superpos_signal}
	r(t) = \sum_{k=1}^K r_k(t-\tau_k)\,,
\end{equation}
where $\tau_k$ denotes the delay of the signal reflected from FBG$_k$.
The columns (atoms) of $\mathbf{A}_k$ are formed by a sampled and delayed version of $r_k(t)$, i.e.
\begin{equation}
\label{eq:dict_atoms}
  [\mathbf{a}_n]_l = r_k(lT_{\text{des}} - n\delta t )  \ ,\quad n\in\Omega_k ,\quad l=1,\dots,L ,
\end{equation}
where $\delta t > 0$ is the delay spacing between the dictionary atoms, $T_{\text{des}}$ is the design 
sampling period used to create the dictionary atoms, and $L$ is the number of samples used to create one dictionary atom. Recall that $L$ is also the number of original signal samples, 
since we create the dictionary atoms based on this signal. 
The value of $T_{\text{des}}$ should be larger or equal to the sampling period of the actual signal acquisition process, $T_s$. Subsequently, we set  $T_{\text{des}} = T_s$.
General dictionaries of the form (\ref{eq:dict_atoms}) are called \emph{shift-invariant}.
In order to cover but not exceed the total observation time, the delay spacing, $\delta t > 0$, has to fulfill $N\delta t  \leq  LT_s  \leq T_{\text{sw}}$.
When the dictionary atoms are created this way, each atom of $\mathbf{A}_k$ is related to a discrete delay in $\mathcal{T}_k$, where 
\begin{equation}
\mathcal{T}_k = \{\tau\in \mathbb{R}_+\, |\, \tau = (n-1)\delta t,\ n\in\Omega_k\}. 
\end{equation}
Finally, the complete composite, shift-invariant dictionary structure for all $K$ FBGs becomes 
\begin{equation}
\label{eq:time-delay-dictionary}
 \mathbf{A} = \left[  \mathbf{A}_{1},\dots,\mathbf{A}_{k},\dots,\mathbf{A}_{K} \right]\,.
\end{equation}
Hence, by identifying the dictionary atoms in $\mathbf{A}$ that most appropriately describe the observed reflections, we directly obtain estimates of the desired reflection delays.

\subsection{Unknown dictionary parameters}
 \label{eq:unknown_params}
Unknown dictionary parameters may arise due to imperfect prior knowledge of the signal and the dictionary.
In this case, the generating functions, $r_k(t)$, depend on an additional set of parameters, $\boldsymbol{\theta}$, i.e.\ $r_k(t,\boldsymbol{\theta}),\ k=1,\dots,K$.
In particular, we introduce uncertain global parameters, $\boldsymbol{\theta}_G$, and/or local parameters, $\boldsymbol{\theta}_{L_k},\ k=1,\dots,K$.
Global parameters account for effects and properties that jointly affect all generating functions $r_k(t,\boldsymbol{\theta}),\ k=1,\dots,K$, while local parameters describe properties of the reflections
from the individual FBGs. 
An example of a global parameter can be the effective receiver bandwidth. Local parameters can be related to the shape of the reflection spectrum of individual FBGs.
Then, an atom associated with $\mathbf{A}_k$ is denoted by $\mathbf{a}_n(\boldsymbol{\theta}_k),\ n\in \Omega_k$, 
where $\boldsymbol{\theta}_k = [\boldsymbol{\theta}_G,\boldsymbol{\theta}_{L_k}]^T $.
Defining $\boldsymbol{\theta} = [\boldsymbol{\theta}_G,\boldsymbol{\theta}_{L_1},\dots,\boldsymbol{\theta}_{L_K}]^T $, the complete parametric dictionary structure is given by  
\begin{equation}
 \label{eq:param_struct_dict}
 \mathbf{A}(\boldsymbol{\theta}) = [\mathbf{A}_1(\boldsymbol{\theta}_1), \dots, \mathbf{A}_k(\boldsymbol{\theta}_k),\dots, \mathbf{A}_K(\boldsymbol{\theta}_K)]   \ .
\end{equation}
Consider the following example with one scalar global parameter and one scalar local parameter, i.e. $\boldsymbol{\theta} = [\theta_G, \theta_{L_1},\dots,\theta_{L_K}]$: 
The overall sensor signal is a superposition of the individual reflections from all FBGs, as in (\ref{eq:superpos_signal}),
where $ r_k(t, \theta_G, \theta_{L_k}),\ k=1,\dots,K$, are now parametrized functions, say  Gaussian functions. 
Their means signify the reflection delays and their variances determine the temporal width of the reflection signals. Now suppose that the amplitudes correspond to the signal attenuation
upon reflection at the FBGs. Then, the individual amplitudes of the reflections can be described by a (local) scalar parameter, $\theta_{L_k}$, in each generating function.
Further assume that all Gaussian functions have the same variance, i.e. the photodetected reflection signals from all FBGs have the same temporal width. %
After passing the receiver circuitry, the widths of the observed reflections are jointly affected by the receiver bandwidth, 
which is described in terms of a (global) scalar parameter, $\theta_G$.  
Based on these considerations, the dictionary atoms of $\mathbf{A}_k(\theta_G, \theta_{L_k}), k=1,\dots,K$, are created as in (\ref{eq:dict_atoms}), i.e.
\begin{equation}
  [\mathbf{a}_n]_l(\theta_G, \theta_{L_k})\ =\ r_k(lT_{\text{des}} - n\delta t,\ \theta_G, \theta_{L_k} )\,, 
\end{equation}
with $ l=1,\dots,L$,\ $n\in\Omega_k$.
Note that we may drop the parameters in the argument but include them when necessary.

\section{Compressed Sampling for Fiber Sensing}
\label{sec:CFS}
Many smart structures need to be constantly monitored, requiring many fiber interrogation cycles. %
This results in a large number of samples to be stored and processed.
Therefore, it is desirable to reduce the number of samples per cycle using CS.

\subsection{Fundamentals of compressed sampling} %
\label{sec:cs_theory}
The sensor signal, $r(t)$, is subsequently described as a vector, \mbox{$\mathbf{r} \in \mathbb{C}^{L}$}, where $L$ is the number of samples \mbox{(without CS).}  %
The fundamental assumption in CS is that, for a given signal, $\mathbf{r}$, there exists a dictionary matrix, $\mathbf{A} \in \mathbb{C}^{L\times N},\ N \geq L$, 
with atoms $\{\mathbf{a}_n\}_{n=1}^N$, in which $\mathbf{r}$ is sparsely \mbox{represented \cite{Baraniuk2007}:} 
\begin{equation}
\label{eq:signalRepresent_r}
\mathbf{r} = \sum_{n=1}^{N} x_n\mathbf{a}_n = \mathbf{A}\mathbf{x}\ .
\end{equation}  
Herein, $N$ is the number of dictionary atoms (discrete reflection delays), and $\mathbf{x} \in \mathbb{C}^{N}$ is a sparse coefficient vector in which all but $\tilde{K}$ entries are zero. 
Then, $\mathbf{r}$ is said to be $\tilde{K}$-sparse in $\mathbf{A}$~\cite{Baraniuk2007}.
Recall that, in the context of the considered application of fiber sensing, the sparsity level, $\tilde{K}$, corresponds to the number of FBGs, $K$. Therefore, we set $\tilde{K}=K$.
The non-zero coefficients in $\mathbf{x}$ indicate the dictionary columns of $\mathbf{A}$, thus, yielding estimates of the reflection delays.\\
It is often encountered in practice that $\mathbf{x}$ is not exactly sparse but contains no more than $\tilde{K}$ significant non-zero \mbox{components \cite{Foucart2013}.} 
Then, $\mathbf{r}$ is said to be \emph{compressible} with respect to $\mathbf{A}$, and can be approximated by sparse reconstruction techniques. %
CS-based acquisition can be described by a sampling matrix $\boldsymbol{\Phi}$, with rows \mbox{$\{\boldsymbol{\phi}_m^T\}_{m=1}^M$ that define $M < N$ projections, i.e. \cite{Baraniuk2007}} 
\begin{equation}
\label{Eq:CSwithbasis}
\mathbf{y} = \sum_{m=1}^{M}\boldsymbol{\phi}_m^T \mathbf{r} + \mathbf{n} = \boldsymbol{\Phi}\mathbf{A}\mathbf{x} + \mathbf{n}\ :=\ \mathbf{Bx} + \mathbf{n}\ . 
\end{equation}
Herein, $\mathbf{B}:=\boldsymbol{\Phi}\mathbf{A} \in \mathbb{C}^{M\times N}$ is the combined sensing matrix, $\mathbf{y}\in\mathbb{C}^{M}$ contains the compressed samples (CS samples),
and $\mathbf{n}$ is the noise component, which is assumed to be zero-mean white Gaussian with variance $\sigma_n^2$. 
Since the data is real-valued, we only deal with real-valued vectors and matrices. 
During each laser sweep, we can acquire the samples described by a single row in $\boldsymbol{\Phi}$ (one projection with $\mathbf{r}$). 
Since $\mathbf{r}$ describes the received sensor signal of one laser sweep, $M$ sweeps are required to complete the total acquisition process.
Moreover, the observed scene is assumed to be stationary during the acquisition time which is determined by $M/S_{\text{r}}$.  
Since $\ M<L$, the number of samples to be stored and processed is reduced. %
Several parallel ADCs can be used to achieve a lower sampling rate in uniform \mbox{sampling  \cite{Strohmer2007}.} 
Non-uniform sampling, described by a sparse sampling matrix, is used to reduce the average sampling rate.

\subsection{Sparse estimation: Methods and guarantees} 
\label{sec:sparse_sig_rec}
In order to obtain a solution for $\mathbf{x}$, we have to solve the equation system in (\ref{Eq:CSwithbasis}). 
However, since $M < N$, the problem is ill-posed and there exist infinitely many solutions. A unique solution can be obtained by imposing appropriate constraints. 
As $\mathbf{x}$ is sparse by assumption, sparse solutions are clearly favorable. 
They can be obtained by greedy or optimization-based methods. 
Optimization-based methods such as \mbox{$\ell_1$-minimization \cite{Foucart2013}} cast (\ref{Eq:CSwithbasis}) as a (usually convex) optimization problem with sparsity constraints but their complexity can be high.
We employ a greedy OMP-based approach, where sparsity is imposed by iteratively choosing one atom that exhibits 
maximum correlation with the residual signal.\\
The success of sparse reconstruction depends on certain joint conditions on $\mathbf{B}$, $\tilde{K}$, and $\mathbf{x}$. 
A popular quality measure is the \emph{restricted isometry property (RIP)} \cite{Candes2005,Candes2008a,Foucart2013}. 
In the context of CS, it can be shown that sub-Gaussian random matrices fulfill the RIP at a given sparsity level with high \mbox{probability \cite{Baraniuk2007,Foucart2013}.}
However, when a redundant dictionary is used, $\mathbf{B}=\boldsymbol{\Phi}\mathbf{A}$ may no longer meet the necessary requirements to guarantee stable and robust sparse reconstruction 
\cite{Donoho2003,Foucart2013,Candes2011}.  
Since $\boldsymbol{\Phi}$ is a random matrix, let us assume that it fulfills the RIP with respect to $\mathbf{A}$ and neglect the aspect of CS for now.
Hence, we regard the problem of estimating a sufficiently sparse representation of the signal from the redundant dictionary $\mathbf{A}$, such that $\boldsymbol{\Phi}$
fulfills the RIP with respect to $\mathbf{x}$. 
In particular, we require that the support $\mathcal{S}=\text{supp}\{\mathbf{x}\}$ is correctly identified since it represents the reflection delays to be estimated.  
In fact, verifying the RIP for a given $\mathbf{A}$ is NP-hard, so alternative coherence measures have been proposed, which are easier to calculate in practice.

\subsubsection{The mutual coherence}
The mutual coherence is a widely used measure to quantify the similarity between dictionary atoms \cite{Donoho2001,Tropp2004,Foucart2013}. It is given by  
\begin{equation}
  \mu(\mathbf{A}) = \max_{j \neq i} | \mathbf{a}_j^H\mathbf{a}_i | = \max_{j \neq i}|[\mathbf{A}^H\mathbf{A}]_{ji}|\ ,\quad j,i \in \Omega\ .
\end{equation}
and reflects the most significant correlation between all pairs. %
\subsubsection{The Babel function}
A more generalized incoherence measure, known as the Babel function or \emph{$\ell_1$-coherence},
is introduced in \cite{Donoho2003,Tropp2004} and \mbox{defined by} %
\begin{equation}
 \label{eq:BABEL}
    \mu_B(\mathbf{A},s) = \max_{\hspace{-0.1cm}|\Lambda|=s}\,\max_{i\in\Omega\setminus\Lambda} \sum_{j\in\Lambda} |\mathbf{a}_i^H\mathbf{a}_j| \; \leq \; s\, \mu(\mathbf{A}) \ .
\end{equation}
It takes into account the dictionary subset $\Lambda \subset \Omega$, with cardinality $|\Lambda|=s$, that exhibits the maximum cumulative correlation with all other dictionary 
atoms $\mathbf{a}_i,\ i \in \Omega \setminus \Lambda$.  
The function $\mu_B(\mathbf{A},s)$ is non-decreasing in $s$, where $\mu_B(\mathbf{A},1) = \mu(\mathbf{A})$. A dictionary is called \emph{quasi-incoherent} if the Babel function increases 
``slowly'' \mbox{in $s$ \cite{Tropp2004}.} 
Using the \emph{exact} recovery condition \cite{Tropp2004}, OMP recovers an ideally $\tilde{K}$-sparse signal if
\begin{equation}
 \mu_B(\mathbf{A},\tilde{K}-1) + \mu_B(\mathbf{A},\tilde{K}) < 1\ .
\end{equation}
For more general signals, however, a good $\tilde{K}$-sparse \emph{approximation} is desired.
The authors in \cite{Tropp2004} show that, if the dictionary is sufficiently incoherent, OMP is an efficient approximation algorithm for finding 
a good $\tilde{K}$-sparse representation which is close to the best $\tilde{K}$-sparse representation of a signal. %

\subsubsection{The coherence distance}
\begin{figure}[t]
  \centering
  \includegraphics[width = 0.80\columnwidth]{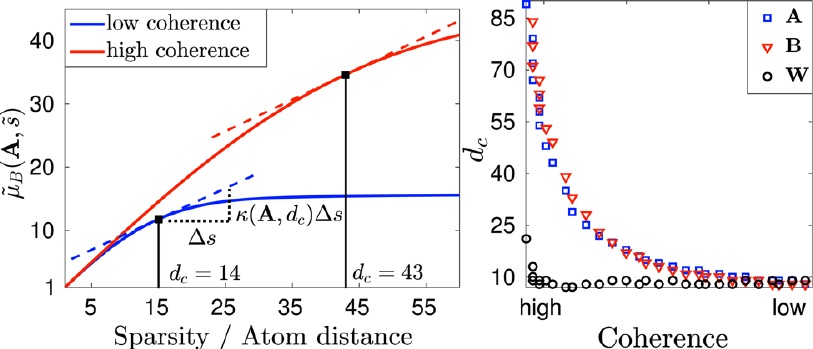}%
  \caption{$\tilde{\mu}_B(\mathbf{A},\tilde{s})$ and $d_c(\mathbf{A})$ for high and low coherence levels (left) and $d_c$ as a function of the coherence level for different matrices: 
  $\mathbf{A}$, $\mathbf{B}$ and $\mathbf{W}$ (right).}
  \label{fig:FBG_simSnapshot}
\end{figure}

Considering the dictionary in (\ref{eq:param_struct_dict}), a strong overlap of adjacent dictionary atoms may result from a small $\delta t$ or from an unlucky 
parametrization, $\boldsymbol{\theta}$. 
In this case, and especially when $\tilde{K}$ is small, $\mu_B(\mathbf{A}(\boldsymbol{\theta}),s)$ and $\mu(\mathbf{A}(\boldsymbol{\theta}))$ do not change much for small variations in $\delta t$ or 
$\boldsymbol{\theta}$.
Yet, this can have strong impact on the level of difficulty in finding the correct sparse support. %
Therefore, to evaluate the level of difficulty of a sparse estimation problem for different parameter settings, 
it is desirable to have an alternative coherence measure that emphasizes this issue. The coherence distance is an appropriate measure for general dictionaries of the 
form (\ref{eq:param_struct_dict}), where the similarity between atoms decreases with increasing index difference. 
Given such dictionary, $\mathbf{A}$, and some $\beta_{\text{cd}} > 0$, we define the coherence distance,
\begin{equation}
  d_c(\mathbf{A},\beta_{\text{cd}}) =  \max_{i\in\Omega}  \left\{  \text{arg}\hspace{-0.0cm}\min_{\hspace{-0.45cm}d = |i-j|}\ \  |\mathbf{a}_i^T\mathbf{a}_j|  \leq \beta_{\text{cd}}\,|\mathbf{a}_i^T\mathbf{a}_i| \  ,\  j\in\Omega \right\}\ ,
\end{equation}
where $|\mathbf{a}_i^T\mathbf{a}_i|\! =\! 1\ \forall\ i \!\in\! \Omega$ by convention.   
$d_c(\mathbf{A},\beta_{\text{cd}})$ corresponds to the smallest difference of atom indices for which the coherence decreases by a factor $\beta_{\text{cd}}$ with respect to the 
atom self-coherence. %
Let us calculate \mbox{$\mu_B(\mathbf{A},s+1)$} and establish a relation to $d_c(\mathbf{A})$. 
For a structured dictionary as \mbox{ in (\ref{eq:param_struct_dict}),} we choose $\Lambda$ in (\ref{eq:BABEL})  \mbox{to be a set of $s+1$} adjacent indices, 
i.e. \mbox{$\Lambda = \{q\in\Omega\, |\, q_0 \leq q \leq q_0 + s \}$.} 
Choosing any \mbox{$i^* \in \Omega\setminus\Lambda$} directly adjacent to $\Lambda$, i.e. \mbox{$|i^*-q_0|=1$ or $|q_0+s - i^*|= 1$,} maximizes the expression \mbox{in (\ref{eq:BABEL})} 
for any \mbox{$q_0\in\{2,\dots,N-s-1\}$.} 
Subsequently, we set $i^* \! = \! q_0 \!-\! 1$. 
Next, we define $\tilde{\mu}_B(\mathbf{A},\tilde{s})$ as a continuous extension of $\mu_B(\mathbf{A},s)$, where $\tilde{s}\in\mathbb{R}_+$ is the continuous variable corresponding to $s\in\mathbb{N}$.
Then, we obtain the \mbox{approximation} %
\begin{equation}
 \mu_B(\mathbf{A},s+1) = \sum_{j=1}^{s+1}|\mathbf{a}_{i^*}^T\mathbf{a}_{i^*+j}| \approx \mu_B(\mathbf{A},s) + \kappa(\mathbf{A},s) \Delta s\ ,    %
\end{equation}
where 
\begin{equation}
  \kappa(\mathbf{A},s) = \frac{d\tilde{\mu}_B}{d\tilde{s}}(\mathbf{A},\tilde{s})|_{\tilde{s}=s}\ .
\end{equation}
Setting $\Delta s = 1$, we find that $\kappa(\mathbf{A},s)\Delta s \geq |\mathbf{a}_{i^*}^T\mathbf{a}_{i^*+s+1}|$ since $\mu_B(\mathbf{A},s)$ is concave,
and obtain the immediate correspondence %
\begin{equation}
  d_c(\mathbf{A},\beta_{\text{cd}}) =\ \ \text{arg} \min_{\hspace{-0.7cm}d\,\in\,\Omega}  \left\{\kappa(\mathbf{A},d) \leq \beta_{\text{cd}}|\mathbf{a}_{i^*}^T\mathbf{a}_{i^*}|\ ,\ i^*\hspace{-0.12cm}+\hspace{-0.03cm} d \leq N\right\}.
\end{equation}
Thus, with the above assumptions on the dictionary structure, $d_c(\mathbf{A},\beta_{\text{cd}})$ is equivalent to the sparsity level $s$, where $\tilde{\mu}_B(\mathbf{A},\tilde{s})$ grows at a 
rate $\kappa(\mathbf{A},d_c) \leq |\mathbf{a}_{i^*}^T\mathbf{a}_{i^*}|/\beta_{\text{cd}} < \kappa(\mathbf{A},d_c-1)$\,.
Subsequently, we always use $\beta_{\text{cd}} = 0.5$ and drop the arguments of $d_c(\mathbf{A},\beta_{\text{cd}})$ when the context is clear. 
For $\beta_{\text{cd}}=0.5$, $d_c$ signifies the index difference at which the atom coherence is decreased by \mbox{3 dB} with respect to the atom self-coherence.
\mbox{Fig.\ \ref{fig:FBG_simSnapshot} (left)} depicts the evolution of $\tilde{\mu}_B(\mathbf{A},\tilde{s})$ in $\tilde{s}$ and its relation to \mbox{$d_c(\mathbf{A},\beta_{\text{cd}}=0.5)$} 
when $\mathbf{A}$ exhibits strong and low coherence, respectively. 
As stated before, the slope of $\tilde{\mu}_B(\mathbf{A},\tilde{s})$ is very similar for both coherence levels when $\tilde{s} < 10$, whereas $d_c$ takes very different values. 
Thus, when $K$ is small, $d_c$ emphasizes the difference in the coherence level of a dictionary for varying \mbox{parametrizations, $\boldsymbol{\theta}$.}
It can be used to quantify the level of difficulty of estimating \mbox{$\mathcal{S} = \text{supp}\{\mathbf{x}\}$.} 
Fig.\ \ref{fig:FBG_simSnapshot} (right) shows the values of $d_c$ at various coherence levels and for different matrices: 
$\mathbf{B}$ inherits much from the structure of the original 
dictionary $\mathbf{A}$. Therefore, their coherence distance takes similar values. $\mathbf{W}$ is a modified dictionary after IAI mitigation (see Section \ref{sec:paramDictLearn}).  
$d_c(\mathbf{W})$ takes significantly lower values which eases the sparse estimation process.

\section{The CS-DL Framework} 
\label{sec:paramDictLearn}
 In this section we introduce the CS-DL framework for WDM-based quasi-distributed fiber sensing. It combines CS-based acquisition with sparse estimation and DL using strongly coherent
parametric dictionaries with uncertainty. 
For CS, a sub-Gaussian matrix can be employed. Sparse binary matrices allow for a simple hardware realization and reduce the average sampling rate. 
At the core of CS-DL, we introduce an algorithm that uses the CS samples to iteratively estimate 
the reflection delays, $\mathcal{S}=\text{supp}\{\mathbf{x}\}$, and the unknown parameters $\boldsymbol{\theta}$. 
Finally, the computational complexity is analyzed.

\subsection{AM-based sparse estimation and DL} 
For sparse estimation and DL, we introduce an AM-based algorithm, called \emph{PDL-OIAI}, which is designed to iteratively estimate a sparse solution, $\mathbf{x}$, in (\ref{Eq:CSwithbasis}), 
along with unknown dictionary parameters, $\boldsymbol{\theta}$, according to Section \ref{sec:working_principle}.\ref{eq:unknown_params}. 
PDL-OIAI is also able to overcome the difficulties of strong dictionary coherence in sparse estimation. %
Unlike related methods in \cite{Raja2016,Austin2013}, PDL-OIAI incorporates IAI mitigation in the AM-based estimation procedure to yield 
a modified sensing dictionary with reduced atom coherence. 
This permits us to use a simple OMP-based algorithm with low complexity for the actual sparse estimation task.   
While different sparse estimation algorithms can be used, such comparison is out of the scope of this work.
Our reasons for choosing OMP are threefold: Firstly, as a greedy method, it is simple and usually faster than optimization-based methods such as 
$\ell_1$-minimization \cite{Cai2011,Foucart2013}. %
Note that algorithms for $\ell_1$-minimization, e.g. LARS, feature relaxed RIP conditions but due to the strong dictionary coherence in our problem, dictionary pre-processing is still necessary and essential to obtain 
a correct sparse solution. 
Secondly, the number of iterations equals the sparsity level, $\tilde{K}$, of the \mbox{signal \cite{Zhang2011,Foucart2013}.} 
Recall that, for the dictionary in (\ref{eq:param_struct_dict}), $\tilde{K}$ also corresponds to the number of reflections, 
i.e. the number of FBGs, $K$, which is exactly known. 
Thirdly, since the reflection delays are continuous, they may fall in between two grid points defined by the dictionary. Optimization-based method tempt to 
compensate this by selecting two adjacent atoms with reduced amplitudes, which misleadingly indicates two reflections. %
OMP only selects one atom in each iteration and guarantees a $\tilde{K}$-sparse approximation after $\tilde{K}$ iterations.
We also define a reference method, called \emph{PDL-OMP}, which does the same as PDL-OIAI but without IAI mitigation.

\subsection{The PDL-OIAI/OMP algorithms}
\label{subsec:pdl-oiai}
\begin{figure}[t!]
\centering
\includegraphics[width=0.6\columnwidth]{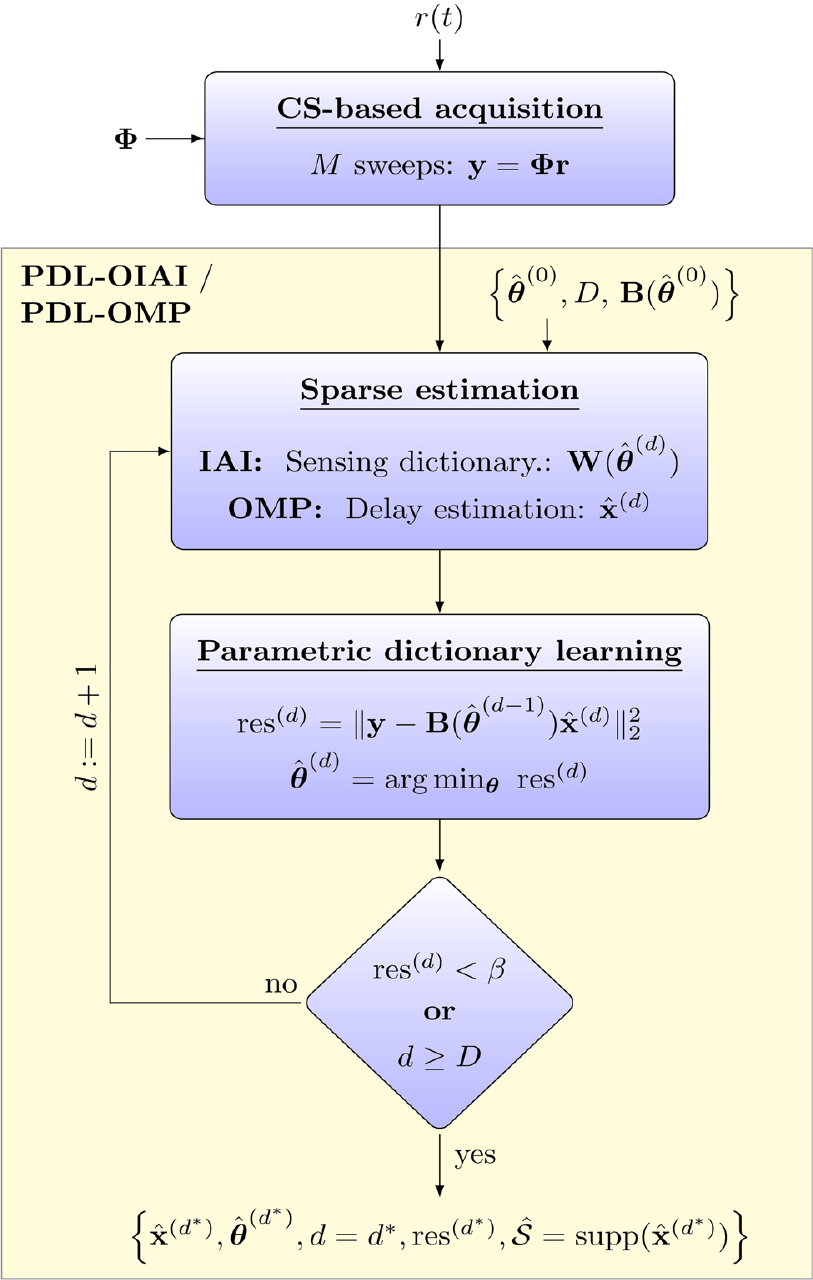}
\caption{Structure diagram of the CS-DL framework: CS-based acquisition followed by PDL-OIAI/OMP for AM-based sparse estimation and PDL. 
PDL-OMP does not perform IAI.}  
\label{fig:parametric_algo}
\end{figure}

Subsequently, we describe the algorithmic details of PDL-OIAI, which are also depicted in the flow-diagram in Fig.\ \ref{fig:parametric_algo}.
The algorithm is initialized using a set of CS samples, $\mathbf{y}$, a maximum number of DL iterations, $D$, 
and the combined matrix, $\mathbf{B}=\boldsymbol{\Phi}\mathbf{A}$, with an initial guess 
of the dictionary \mbox{parameters, $\hat{\boldsymbol{\theta}}\,^{(0)}$.} Some prior information of the parameters is usually available, so that an initial value can be chosen based on the 
experience and expertise of the operator or based on physical constraints.
PDL-OIAI adopts the IAI mitigation method in \cite{Yang2010} to reduce the coherence distance of $\mathbf{B}$. 
A brief review of this method is given in Subsection \ref{subsubsec:IAI}. 
It yields a modified \emph{sensing dictionary}, $\mathbf{W}$,
which maintains the physical interpretation of the sparse support (i.e. the reflection delays) with respect to $\mathbf{A}$. Thus, the original sensor signal can be 
estimated by $\mathbf{A}\hat{\mathbf{x}}$, where $\hat{\mathbf{x}}$ is estimated using $\mathbf{W}$, 
and $\hat{\mathcal{S}}:=\text{supp}\{\hat{\mathbf{x}}\}$ contains the appropriate delay estimates.
Recall that PDL-OMP does not perform IAI-mitigation.
Subsequently, $\boldsymbol{\theta}$ is estimated by minimizing  
the residual, $\text{res} := \text{res}(\boldsymbol{\theta})$. 
In \mbox{Fig.\ \ref{fig:parametric_algo}} we use the $\ell_2$-norm to calculate the residual, but any other suitable objective function 
may be used. The pre-defined range of possible parameter values %
can be initially defined according to prior knowledge and physical limitations. This scheme yields locally optimal parameter estimates as it cannot be 
guaranteed that the parameter ranges form a convex set, and also the objective function of the residual may not be jointly convex in all 
uncertain parameters. 
In the $d$-th iteration, the estimates $\hat{\boldsymbol{\theta}}\,^{(d)}$ are used to obtain an improved estimate, $\hat{\mathbf{x}}\,^{(d+1)}$,
which, in turn, yields an improved estimate $\hat{\boldsymbol{\theta}}\,^{(d+1)}$. 
The algorithm terminates when $\text{res}^{(d)}$ is below a threshold or after $d=D$ iterations. The threshold can be set to $\beta = (1+\epsilon_r)P_n$ 
with $\epsilon_r > 0$, where $P_n$ is an estimate of the total noise power. %
Ultimately, the iteration index $d=d^*$, associated with the smallest residual, is determined to output the final solutions, $\hat{\mathbf{x}}\,^{(d^*)},\hat{\boldsymbol{\theta}}\,^{(d^*)}$, and $\hat{\mathcal{S}} := \text{supp}\{\hat{\mathbf{x}}^{(d^*)}\}$.

\subsection{IAI mitigation} 
\label{subsubsec:IAI}
We briefly review the concept of IAI mitigation in \cite{Yang2010}. It is adopted in PDL-OIAI to ease the task of sparse estimation from highly coherent 
dictionaries. 
In contrast to the methods \mbox{in \cite{Elad2007,Schnass2008}}, it also takes the received data into account.
The coherence of the combined matrix $\mathbf{B}$ originates from the coherence of $\mathbf{A}$ (Fig.\ \ref{fig:FBG_simSnapshot} (right)). 
Therefore, IAI mitigation is applied to $\mathbf{B}$ directly to yield a modified sensing dictionary. 
OMP suffers from IAI, which lies in its conceptual approach to sparse estimation.
OMP attempts to iteratively identify the atoms in $\mathbf{B}=[\mathbf{b}_1,\dots,\mathbf{b}_N]$ that minimize the residual signal. In the \emph{first} 
iteration ($k=1$), an atom is selected by \mbox{solving \cite{Yang2010}}
\begin{equation}
\label{eq:OMP}
  \max_{i \in \Omega} |\mathbf{b}_i^T\mathbf{y}|\ =\ \max_{i \in \Omega} \left|\sum_{q\in\mathcal{S}}^{} \mathbf{b}_i^T\mathbf{b}_q x_q + \mathbf{b}_i^T\mathbf{n}\right|,  
\end{equation}
where we are interested in determining the non-zero components \mbox{$x_q \neq 0$}. The sum-term in (\ref{eq:OMP}) describes interference with other 
atoms $\mathbf{b}_q,\ q\in\mathcal{S}$, which leads to the selection of incorrect atoms. 
Thus, instead of (\ref{eq:OMP}), Yang \mbox{\textit{et al.} \cite{Yang2010}} suggest to solve 
\begin{equation}
    \max_{i \in \Omega} \left|  \sum_{q\in \mathcal{S}}^{} \mathbf{w}_i^T\mathbf{b}_qx_q + \mathbf{w}_i^T\mathbf{n} \right| \ ,
\end{equation}
where $\mathbf{w}_i,\ i=1,...,N$, are the atoms of the modified sensing dictionary, $\mathbf{W}$. %
For $i\in\mathcal{S}$, the purpose of $\mathbf{w}_i$ is to minimize IAI with other \emph{correct} atoms $\mathbf{b}_j,\ j\in\mathcal{S} \setminus i$, while maintaining the 
correlation with $\mathbf{b}_i$. 
This can be achieved using the minimum interference distortionless response (MIDR) approach in \cite{Yang2010}, i.e. by choosing $\mathbf{w}_i$ such that
\begin{equation}
 \label{eq:IAI_w}
  \min_{\mathbf{w}_i} \left|\mathbf{w}_i^T\mathbf{B}_\mathcal{S}\mathbf{B}_\mathcal{S}^T\mathbf{w}_i\right| \quad \text{s.t.} \quad \mathbf{b}_i^T\mathbf{w}_i = 1\ ,
  \quad i=1,\dots,N\ .
\end{equation}
$\mathbf{B}_{\mathcal{S}}$ contains the columns of the unknown true support. %
Therefore, $\mathbf{B}_\mathcal{S}\mathbf{B}_\mathcal{S}^T$ is iteratively approximated by $\mathbf{B}\mathbf{U}^{(j)}\mathbf{B}^T$, 
where \mbox{$\mathbf{U}^{(j)} = \text{diag}(|(\mathbf{W}^{(j-1)})^T\mathbf{y}|^\rho)$}, $j\in\{1,\dots,J\}$ is an iteration index and $\rho>0$ is a regularization parameter. 
There exists a closed-form solution for each $\mathbf{w}_i\,^{(j)},\ i=1,\dots,N$. These are gradually improved in $J$ internal iterations with initialization $\mathbf{W}^{(0)} = \mathbf{B}$
(see \cite{Yang2010} for details).
In subsequent OMP iterations ($k>1$), $\mathbf{y}$ has to be replaced by the current residual, $\mathbf{g}^{(k)}, k=1,\dots,\tilde{K}=K$, 
where $\mathbf{g}^{(0)} := \mathbf{y}$. 
The constraint in (\ref{eq:IAI_w}) ensures that $\mathcal{S}$ still corresponds to the reflection delays to be estimated. %
That is, $\mathbf{r}$ can be estimated by $\hat{\mathbf{r}} = \mathbf{A}\hat{\mathbf{x}}$, where $\hat{\mathbf{x}}$ is a solution to $\mathbf{y} = \mathbf{Wx}$, 
and $\hat{\mathcal{S}}  = \text{supp}\{\hat{\mathbf{x}}\}$ contains appropriate delay estimates.
As shown in Fig.\ \ref{fig:FBG_simSnapshot} (right), for the sensing dictionary $\mathbf{W}:=\mathbf{W}^{(J)}$, 
we indeed find that $d_c(\mathbf{W}) \ll d_c(\mathbf{B})$. 

\subsection{Computational Complexity}
\label{subsec:compCompl}
PDL-OIAI/OMP performs $D$ AM iterations, i.e. sparse estimation and subsequent PDL.
For sparse estimation using OMP, both algorithms require $\tilde{K}=K$ iterations according to the considered sparsity level of $\mathbf{x}$ (number of FBGs), 
where each iteration has complexity $\mathcal{O}(NM)$. Thus, in the sparse estimation step of the $d$-th AM iteration, $d=1,\dots,D$, PDL-OMP has complexity $\mathcal{O}(KMN)$.
However, the complexity of PDL-OIAI, in the $k$-th OMP iteration,
is dominated by calculating $\mathbf{w}_i$ in (\ref{eq:IAI_w}) $\forall\ i=1,\dots,N$, which has complexity $\mathcal{O}(N^2M+NM^2+M^3)$.
This step is repeated in $J$ internal iterations (usually $J<10$) to improve the quality of $\mathbf{W}$ for IAI mitigation.
Then, in the sparse estimation step of the $d$-th AM iteration, PDL-OIAI has complexity $\mathcal{O}(KJ(N^2M+NM^2+M^3))$.
As stated in \cite{Yang2010}, a significant speedup can be achieved by calculating $\mathbf{W}$ only once for $k=1$, although this results in a 
lower performance. 
Moreover, $\mathbf{w}_i$ can be calculated independently $\forall\ i=1,\dots,N$, which is efficiently implemented using parallel processing.\\
In the $d$-th AM iteration, sparse estimation is followed by PDL. 
A closed-form solution for $\boldsymbol{\theta}$ is not generally available. Given that there are $|\boldsymbol{\theta}| = P$ parameters to be estimated,
the residual can be locally minimized for each $\theta_p, p=1,\dots,P$, by varying its value through a set of possible values, $\mathcal{R}_{\theta_p}$, %
while keeping all other parameters fixed. 
We use such a grid search with a constant spacing, assuming the same number of possible values for each parameter, i.e. $|\mathcal{R}_{\theta_p}| = R_{\theta}\ \forall\  p=1,\dots,P$.
The complexity of computing the residual for one parameter $\theta_p \in \mathcal{R}_{\theta_p}$ with $R_\theta$ grid points is $\mathcal{O}(R_\theta NM^2)$.
Hence, for $P$ parameters, where each one has $R_\theta$ grid points, the complexity is $\mathcal{O}((R_{\theta}NM^2)^P)$. This is the complexity of the PDL step in the $d$-th AM iteration for PDL-OIAI/OMP.\\
\mbox{Table \ref{tb:complexity}} summarizes the overall computational complexity of PDL-OIAI and PDL-OMP.  
\begin{table}[H]  \centering
 \caption{Computational complexity of PDL-OMP and PDL-OIAI when $\mathbf{W}$ is determined \mbox{$\forall\ k=1,\dots,K$,} or fixed for $k>1$.} \vspace{0.3cm}
 \label{tb:complexity}
%
\begin{tabular}{p{2.65cm}p{0.545\textwidth}}   
  \toprule[1.5pt] \\[-1.4ex]
  \hspace{0.3cm}\textnormal{\textbf{PDL-OMP}}: 			 &	\hspace{-0.6cm}	 	$\mathcal{O}\hspace{-0.06cm}\left(  D [\, KNM + (R_{\theta}NM^2)^P\, ] \, \right)$		 \\[0.9ex]%
  \hspace{0.3cm}\textnormal{\textbf{PDL-OIAI}}:  			 &	\hspace{-0.6cm}	 	$\mathcal{O}\hspace{-0.06cm}\left(  D [\,   KJ(N^3M + N^2M^2 + NM^3)	+  	 (R_{\theta}NM^2)^P\,          ] \, \right)$		 		 	\\[0.9ex] %
  \hspace{0.3cm}\textnormal{\textbf{$\rightarrow\hspace{-0.1cm}\mathbf{W}$ fixed:}}  &  \hspace*{-0.7cm}  $\mathcal{O}\hspace{-0.06cm}\left( D [   J(N^3M\hspace{-0.06cm} +\hspace{-0.06cm} N^2M^2\hspace{-0.09cm} +\hspace{-0.06cm} NM^3)\hspace{-0.09cm}	+\hspace{-0.06cm} KNM\hspace{-0.06cm} +\hspace{-0.06cm} (R_{\theta}NM^2)^P         ] \right)$ 	 	\\[0.9ex] %
  \bottomrule[1.5pt]
\end{tabular}  \vspace{-0.03cm}
\end{table}

\section{Performance Evaluation in Simulations and Experimental Validation Using Real Data}
\label{sec:CS_DL_pratctice}
It is time to
apply the CS-DL framework to a particular WDM-based distributed sensing system. 
We compile a parametric model for the received signal to
construct a specific instance of the parametric dictionary in (\ref{eq:param_struct_dict}).
Then, we use simulations to evaluate the performance of PDL-OIAI and PDL-OMP for the considered system in various scenarios of different sampling matrices, dictionary parameter 
values, and SNRs. Finally, we provide some experimental validation by applying CS-DL to real sensor data using the same parametric 
model/dictionary. %

\subsection{A parametric signal model for CS-DL} %
\label{sec:param_model}
To apply CS-DL in practice, a modular system architecture is considered. Extending the work in \cite{Weiss2013}, we adopt and compile existing physical models 
to express the received sensor signal and to construct a specific instance of the parametric dictionary in (\ref{eq:param_struct_dict}).  
Modularity means that each component has a well-defined input/output relation. We further assume that the subsystem between laser and photodetector 
can be described by a linear time-invariant (LTI) system. %
We present a specific model for the basic system architecture in \cite{Nakazaki2009,Yamashita2009}, which is depicted in 
Fig.\ \ref{fig:cfsp}. %
The basic components are described below.

\subsubsection{Laser output:}
The authors in \cite{Nakazaki2009,Yamashita2009} have developed a special class of $\lambda$-tunable lasers. 
Their advantages for distributed fiber sensing are three-fold: firstly, no tunable filters are required, which keeps the hardware complexity low. 
Secondly, a wide tuning bandwidth is available to serve a large number of FBGs. 
Thirdly, high sweep rates are achieved, where the wavelength varies linearly with the frequency of the modulation current. 
The emitted light at the center wavelength $\lambda$ can be described in the frequency domain by a chirped Gaussian pulse of the %
form \cite{Nakazaki2009,Yamashita2009,Tamura1996}    
\begin{equation}
\label{eq:pulse_freq}
A^{(\lambda)}(\omega) \sim \text{exp}\left(-\frac{\omega^2}{2\,(\delta\omega)^2}\right)\ ,
\end{equation}
with a wavelength-dependent bandwidth 
\begin{equation}
\label{eq:pulseBW}
\delta\omega = \sqrt{\pi\frac{f_{m}}{\lambda}}\left(\frac{8\pi c_0\Gamma_a}{|D_{\text{\tiny{DCF}}}\,\hspace{-0.1cm} ^{(\lambda)}|  L_{\text{\tiny{DCF}}} }\right) \ .
\end{equation}
$c_0$ is the speed of light and $L_{\text{\tiny{DCF}}}$ is the length of the dispersion compensating fiber (DCF) which forms the laser cavity. 
$\Gamma_a$ denotes the amplitude modulation index of the light within the DCF and $D_{\text{\tiny{DCF}}}\,\hspace{-0.1cm} ^{(\lambda)}$ is the first order dispersion parameter.
The latter can be determined by a 5th-order Sellmeier fit to the measured group delay \cite{Born1987,Pal2005}.
A semiconductor optical amplifier (SOA) is used as a gain medium and $f_m$ is the modulation frequency of the injection current.
The $i$-th emitted laser pulse is comprised of the superimposed cavity modes around the instantaneous center wavelength, $\lambda_i$.  
At index $i$, the instantaneous pulse repetition rate is given by $\tau_{\text{rep},i} =  1/f_{m,i}$.
The sum of all pulses forms the output signal corresponding to one laser sweep. 
This signal is modulated by the wavelength-dependent gain profile of the SOA. It ise described by~\cite{Willner1995,AbdElAziz2010} %
\begin{equation}
\label{eq:SOA_gain}
\breve{g}_{\text{\tiny SOA}}^{(\lambda)} = \frac{a_1(N_c-N_0) - a_2(\lambda - \lambda_{N_c})^2 + a_3(\lambda - \lambda_{N_c})^3 )}{1 + \epsilon_c P_{\text{av}}}  ,
\end{equation}
where $\epsilon_c$ is called compression factor, 
$P_{\text{av}}$ is the average output power over the length of the SOA,
$N_c$ is the actual carrier density and $N_0$ is the carrier density at the transparency point (no-gain wavelength). In (\ref{eq:SOA_gain}),
$\lambda_N$ and $\lambda_0$ are the maximum gain wavelengths corresponding to $N_c$ and $N_0$, respectively. 
The difference $\lambda_{N_c}-\lambda_0$ is due to the carrier difference $N_c-N_0$, while
$a_1$ accounts for the carrier difference with respect to the transparency point and
$a_2$ characterizes the spectral width of the gain profile with asymmetry indicated by $a_3$. 
Note that $a_{1,2,3}$ can be determined by a 3rd-order fit to the measured SOA gain. In our simulations, we used some of the values listed \mbox{in \cite{Willner1995,AbdElAziz2010}.}

\subsubsection{Fiber transmission:}
The transfer function of a standard single-mode fiber (SMF) of length $L_{\text{\tiny SMF}}$ can be derived from a Taylor expansion of the propagation constant $\beta\,^{(\lambda)}$, where we neglect any 
non-linear effects by assumption. 
Since the FBGs are located at different positions along the sensing fiber, the travel distance varies in $\lambda$, 
i.e. $L_{\text{\tiny SMF}}^{\text{\tiny $(\lambda)$}}$. 
Denoting the damping by $\alpha^{(\lambda)}$ and the 2nd- and 3rd-order terms of the Taylor expansion of $\beta\,^{(\lambda)}$ by $\beta_2\,^{(\lambda)}$ and $\beta_3\,^{(\lambda)}$, respectively,
the baseband transfer function becomes \cite{Savory2008,Seimetz2009,Ip2008} 
\begin{equation}
\label{eq:H_w}
\hspace{-0.1cm}H^{(\lambda)}\hspace{-0.1cm}\left(\omega,L_{\text{\tiny SMF}}^{(\lambda)}\right) = \text{exp}\left(\hspace{-0.1cm}-\left(\alpha^{(\lambda)} + j\frac{\beta_2^{(\lambda)}}{2}\omega^2 + j\frac{\beta_3^{(\lambda)}}{6}\omega^3 \right)L_{\text{\tiny SMF}}^{(\lambda)}\hspace{-0.1cm}\right).  %
\end{equation}
The parameters $\beta_{2}^{(\lambda)}$ and $\beta_{3}^{(\lambda)}$ are related to the dispersion coefficient, $D_{\text{\tiny SMF}}^{\text{\tiny$(\lambda)$}}$, and to the dispersion 
slope, \mbox{$S_{\text{\tiny SMF}}^{\text{\tiny$(\lambda)$}} = (\text{\small $d/d\lambda$})D_{\text{\tiny SMF}}^{\text{\tiny$(\lambda)$}}$}, \mbox{respectively \cite{Seimetz2009}.}
The damping is modeled \mbox{by \cite{Mendez2007}}
\begin{equation}
  \alpha^{(\lambda)} = A_R\frac{1}{\lambda^4} + B + C\,^{(\lambda)}\ .
\end{equation}
$A_R$ denotes the Rayleigh scattering coefficient, $B$  accounts for other wavelength-independent losses such as microbending or waveguide imperfections 
and $C\,^{(\lambda)}$ describes other wavelength-dependent losses such as OH$^-$ absorption peaks.
These parameters can be determined by a least-squares fitting to the measured damping. A model for $C\,^{(\lambda)}$ is 
available \mbox{in \cite{Mendez2007}}.

\subsubsection{FBG reflection:}
We consider uniform (non-chirped) FBGs that are modeled by a periodic perturbation of the refractive index along the fiber core in 
propagation ($z$-) direction \cite{Venghaus2006,Kashyap2009}:
\begin{equation}
n_c(z) = n_0 + \Delta n_c\ \text{cos}\!\left(\frac{2\pi (z-z_0)}{\Lambda_{\text{\tiny FBG}}}\right)\ ,\ z\in[z_0,z_0+L_{\text{\tiny{G}}}]\ ,
\end{equation}
where $L_{\text{\tiny{G}}}$ is the total grating length, $z_0$ is the grating location,
$n_0$ is the average refractive index within one spatial period $\Lambda{\text{\tiny FBG}}$, 
and $\Delta n_c$ is the perturbation 
amplitude. 
Using \mbox{coupled-mode} \mbox{theory \cite{Yariv1973,Kogelnik1988,Venghaus2006,Kashyap2009}}, the spectral characteristics of an FBG can be derived by calculating solutions for the field 
amplitudes, $S(z)$, $R(z)$, of a mode and an identical counter-propagating mode, respectively. Coupling is achieved by the dielectric perturbation.
For a uniform grating, this corresponds to solving a system of two coupled 1st-order ordinary differential equations with 
constant coefficients and appropriate boundary \mbox{conditions \cite{Erdogan1997}}
\begin{equation}
\label{eq:FBG_DGL}
  \frac{dR}{dz}  = i\,\sigma_c R(z) + i\,\kappa_c S(z)\  ; \quad\  \frac{dS}{dz} = -i\,\sigma_cS(z) + i\,\kappa_c^* R(z) \ ,
\end{equation}
which has a closed-form solution. In (\ref{eq:FBG_DGL}), $i$ denotes the imaginary unit and $(\cdot)^*$ the complex conjugate, while
$\sigma_c$ and $\kappa_c$ are called the DC-/AC-coupling coefficients. They are proportional to the refractive index.
The ratio $\rho^{(\lambda)} = \left. S(z=z_0)/R(z=z_0)\right|_{\lambda}$ describes the field amplitude reflection coefficient at wavelength $\lambda$.
Defining $\gamma_c = (\kappa_c^2 - \sigma_c^2)^{1/2}$, it is given by \cite{Erdogan1997}
\begin{equation}
\label{eq:rho}
\rho^{(\lambda)} = \frac{-\kappa_c\;\text{sinh}(\gamma_c\, L_{\text{\tiny{G}}})}
{ \sigma_c\;\text{sinh}(\gamma_c\,L_{\text{\tiny{G}}})  +  i\,\gamma_c\,\text{cosh}(\gamma_c\,L_{\text{\tiny{G}}}) }\ . 
\end{equation}
To model more sophisticated spectral characteristics, e.g. for non-uniform gratings, including chirp or apodization, 
a closed form solution can be derived by adopting a piece-wise 
uniform approach, where the FBG is subdivided into $M_G$ uniform segments. Then, the field amplitudes are calculated for each segment separately. 
The result is given in matrix form~\cite{Erdogan1997}
\begin{equation}
 [R_{M_G}, S_{M_G}]^T =\ \ \mathbf{F}\ [R_0, S_0]^T\ ,
\end{equation}
where $\mathbf{F} = \mathbf{F}_{M_G} \cdot \mathbf{F}_{M_G-1} \cdots \mathbf{F}_1$  is the total transition matrix 
and $\mathbf{F}_i\,,\ i=1,\dots,M_{G}$, are the transition matrices of each segment.

\subsubsection{Received sensor signal:} 
The reflected signal travels the same way back to the receiver. 
Thus, $H^{(\lambda)}(\omega,L_{\text{\tiny SMF}}^{\text{\tiny $(\lambda)$}})$ is applied twice. The optical center frequency of the total sweep range is \mbox{$\Omega_0 = c_0/\lambda_0$} and
the instantaneous center frequency of the $i$-th pulse is \mbox{$\Omega_i = \Omega_0 - \Delta\omega_i$,} where $\Delta\Omega_i = c_0(\lambda_i-\lambda_0)/(\lambda_i\lambda_0)$. %
Using (\ref{eq:pulse_freq},\,\ref{eq:SOA_gain},\,\ref{eq:H_w},\,\ref{eq:rho}) 
we obtain a frequency domain expression in the baseband for the received pulse train reflected from an FBG at distance $L_z$:
\begin{equation}
E_r(\omega) = \sum_i \breve{g}_{\text{\tiny SOA},i}\,\rho_i\, e^{-j\omega\tau_{\text{rep},i}}A_i(\omega-\Delta\Omega_i)\, H_i(\omega-\Delta\Omega_i,L_z)^2.  %
\end{equation}
Index $i$ indicates the dependence of each term on $\lambda_i$. 
The time-domain signal, $E_r(t) = E_r(t,\lambda(t))$, is obtained by taking the inverse Fourier transform of $E_r(\omega)$, where $\lambda(t)$ 
indicates the temporal evolution of the wavelength.
In the photodetection process, the responsitivity of the PD, $R_{\text{\tiny PD }}^{\text{\tiny$(\lambda)$}}$, depends on the quantum 
efficiency of the material, $\eta_{\text{\tiny PD }}^{\text{\tiny$(\lambda)$}}$ \cite{Alexander1997}. Different PD types can be distinguished.
For simplicity, we assume a uniform distribution of the intensity, $I(\mathbf{q},t)$, at all locations $\mathbf{q} \in \mathbb{R}^2$ within the sensitive area, $A_{\text{\tiny PD}}$, of the PD. 
Then, the incident power becomes 
$P(t) = \int_{A_{\text{\tiny PD}}} I(\mathbf{q},t) dA_{\text{\tiny PD}} = |E_r(t)|^2/Z_W$, where $Z_W$ is the wave impedance of the medium \cite{Alexander1997}. 
The photocurrent is \cite{Alexander1997}
\begin{equation}
\label{eq:I}
i(t) \; = R_{\text{\tiny PD }}^{\text{\tiny$(\lambda)$}} P(t) = \frac{q_e\,\lambda(t)\,\eta_{\text{\tiny PD }}^{\text{\tiny$(\lambda(t))$}}}{h\,c_0}\frac{|E_r(t)|^2}{Z_W} \ , 
\end{equation}
where $h$ is Planck's constant and $q_e$ is the elementary charge \cite{Alexander1997}. 
A slow photodiode can demodulate the pulse train envelope.  %
To this end, we model the effective bandwidth of the receiver circuitry, $\Delta f$, by a lowpass filter with transfer function $H_{\text{\tiny{LP}}}(\omega)$: 
\begin{equation}
\label{eq:i_t}
r(t,\Delta f) = \frac{1}{2\pi}\int_{-\infty}^{\infty}e^{j\omega t}\, \left[H_{\text{\tiny{LP}}}(\omega) i(\omega)\right] \, d\omega\ ,
\end{equation}
where $i(\omega)$ is the Fourier transform of $i(t)$. 
In the ensuing sections, we assume uncertainty in $\Delta f$ as a global parameter, $\theta_G$, 
which jointly affects the temporal width of the reflections contained in $r(t,\Delta f)$.
Ultimately, a parametric dictionary, $\mathbf{A}$, can be created from $r(t,\Delta f)$
according to (\ref{eq:param_struct_dict}).

\subsection{Simulations} 
\label{sec:sim_results}
In order to assess the performance of sparse estimation and DL, we have truncated the CS-DL framework to suit the particular architecture in Fig.\ \ref{fig:cfsp}. 
The basic model parameters have been tuned to match the system in \cite{Nakazaki2009,Yamashita2009}. Recall that $r(t)$ represents 
the generating function of the dictionary according to (\ref{eq:param_struct_dict}). 
Also recall, that our main goal is to estimate the reflection delays (i.e. the underlying sparse representation) based on the highly redundant combined sensing dictionary, $\mathbf{B}$.
Since sparse estimation methods are computationally complex, we apply CS not only to reduce the sampling rate but also the number of samples to be processed. 
However, a smaller number of measurements, $M$, increases the dictionary coherence and lowers the expected estimation accuracy. %

\subsubsection{Scenario and basic settings}
We consider a sensing fiber with $K=3$ FBGs. Their Bragg wavelengths are such that two of the temporal reflections are closely spaced. 
All reflections have a uniform shape, a common amplitude, $A_x$, and a temporal width determined by the true value of $\Delta f$.
As a global dictionary parameter, we estimate $\Delta f$ in terms of a normalized parameter $\hat{\theta} = \widehat{\Delta f} / \Delta f$.
An initial value is chosen at random with $\hat{\theta}\,^{(0)} \in [1.2, 5]$ and $R_\theta < 100$. %
The original \mbox{signal, $\mathbf{r}$,} has length $L = 134$ and the dictionary has $N = 2L$ atoms with a delay resolution of $\delta t = 50$ ns.
CS samples are taken by three types of sampling matrices, $\boldsymbol{\Phi}$, with i.i.d. entries drawn from the following random distributions:\\[0.00cm]
\begin{center}
\begin{tabular}{lll}
  (a) Gauss :& \hspace{-0.3cm}$\mathcal{N}(0,1)$ & \\[0.2cm]
  (b) Rademacher :& \hspace{-0.3cm}$\{\pm 1\}$ &\hspace{-0.3cm}with equal probability\\[0.2cm]
  (c) DF \cite{Achlioptas2003}: & \hspace{-0.31cm}$\{-1,0,1\}$ &\hspace{-0.3cm}with probabilities $\{\frac{1}{6},\frac{2}{3}, \frac{1}{6}\}$\ .
\end{tabular}\\[0.4cm]
\end{center}
PDL-OIAI and PDL-OMP run 3 OMP-iterations to identify \mbox{$\tilde{K}=K=3$} atoms corresponding to the reflections. 
PDL-OIAI performs additional $J=10$ internal iterations for IAI mitigation. The regularization parameter for IAI mitigation was set to $\rho=1$.
The sensing dictionary, $\mathbf{W}$, is only calculated once in the first OMP iteration to speed up computations 
(see \mbox{Section \ref{sec:paramDictLearn} \ref{subsec:compCompl})}. %
We set the maximum number of AM iterations to $D = 8$. 
For our simulations, we used an Intel$^{\text{\tiny \textregistered}}$ Core$^{\text{\tiny \texttrademark}}$ i5 CPU 760 with 4 cores.
The average computation time per AM iteration, $d$, was approximately $0.8$\,s  for PDL-OMP, and $1.4$\,s for PDL-OIAI.\\     
We evaluate the performance in terms of the root mean-squared error (RMSE). 
For a vector $\mathbf{v}$ and its estimate $\hat{\mathbf{v}}$, we define
\begin{equation}
 \text{RMSE}(\mathbf{v},\hat{\mathbf{v}}) = \sqrt{E\left\{\|\mathbf{v}-\hat{\mathbf{v}}\|_2^2\right\}}\ ,
\end{equation} 
where the expectation operator is replaced by the sample mean over Monte Carlo runs to yield an approximation termed $\overline{\text{RMSE}}$.
Further, we calculate the Cram{\'e}r-Rao lower bound (CRLB) for sparse estimation. %
It serves as a benchmark to compare the performance of PDL-OIAI with achievable theoretical bounds.
The constrained CRLB describes a lower bound for the variance of any unbiased estimator, $\hat{\mathbf{x}}_{\text{ub}}$, in a sparse setting \mbox{with $\tilde{K}=K$ \cite{Ben-Haim2010}:}
\begin{equation}
 \label{eq:constr_CRLB}
  \text{CRLB:}\quad E\{\|\hat{\mathbf{x}}_{\text{ub}} - \mathbf{x}\|_2 ^2\} \geq \sigma_n^2\, \text{Tr}\left( [\mathbf{B}_{\mathcal{S}}^T \mathbf{B}_{\mathcal{S}}]^{-1} \right),\ \|x\|_0 = K.
\end{equation}
Herein, $\mathbf{B}_{\mathcal{S}}$ is a submatrix of $\mathbf{B}$ that consists of only the columns with indices in $\mathcal{S}\hspace{-0.08cm} := \text{supp}\{\mathbf{x}\}$
and Tr\,($\cdot$) denotes the trace operator on a matrix.
The lowest value of the CRLB, \mbox{i.e. $K\,\sigma_n^2$,} is achieved when the columns in $\mathbf{B}$ are orthogonal \mbox{(see Appendix \ref{app:A})}.
It can be shown \cite{Ben-Haim2010} that the constrained CRLB is equal to the CRLB of the \emph{oracle estimator}, which has perfect knowledge of $\mathcal{S}$.
\begin{figure}[t]
  \centering
  \includegraphics[width = 0.7\columnwidth]{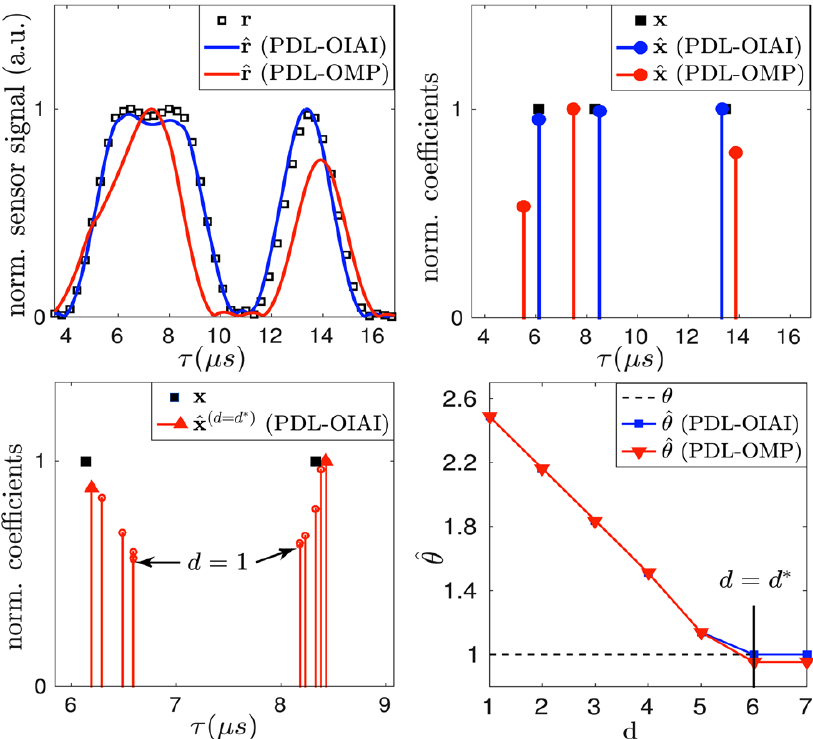}\vspace{0.3cm}
  \caption{Simulation results for $d_c = 14$: noiseless sensor signal (top left), sparse signal (top right). Black bullets indicate the true delays. 
                Evolution of sparse coefficient estimates (zoomed) vs. iteration index $d$ (bottom left), and $\hat{\theta}$ vs. $d$ (bottom right).} 
  \label{fig:sim_realization}
\end{figure}

\subsubsection{Visualization of PDL-OIAI/OMP: Iterative estimation progress}
Simulated sensor data is created with an SNR of 10 dB. 
We take $M=94$ CS samples using a Gaussian sampling matrix, which reduces the number of samples by $30\%$.
The coherence distance of the dictionary is $d_c = 14$. 
Fig.\ \ref{fig:sim_realization} (top left) shows the original samples of the noiseless sensor signal, $\mathbf{r}$, and the estimates, $\hat{\mathbf{r}}$, 
obtained by PDL-OIAI/OMP.
The sparse signal estimates are depicted in Fig.\ \ref{fig:sim_realization} (top right) and their support indicates the reflection delays. 
It can be seen that PDL-OMP cannot deal with highly coherent atoms and chooses the wrong support, while PDL-OIAI accurately identifies the true 
support. 
For PDL-OIAI, the sparse support (delay estimates) of the two closely spaced sources are shown in different iteration steps in Fig.\ \ref{fig:sim_realization} (bottom left). 
We can observe how the estimates for support and amplitudes of $\mathbf{x}$ improve with increasing iteration index, $d$.
\mbox{Fig.\ \ref{fig:sim_realization}} (bottom right) shows the evolution of $\hat{\theta}\,^{(d)}$ vs. $d$.
Due to a limited step size per iteration, the slope is linear for $d<5$ as both algorithms choose the maximum step-size towards the correct value.
PDL-OIAI and PDL-OMP both converge to a final value $\hat{\theta} \approx \theta = 1$ after $d^* = 6$ iterations.

\newpage

\subsubsection{Performance evaluation}
\begin{figure}[H]
\centering\vspace{2cm}
\includegraphics[width = 0.75\columnwidth]{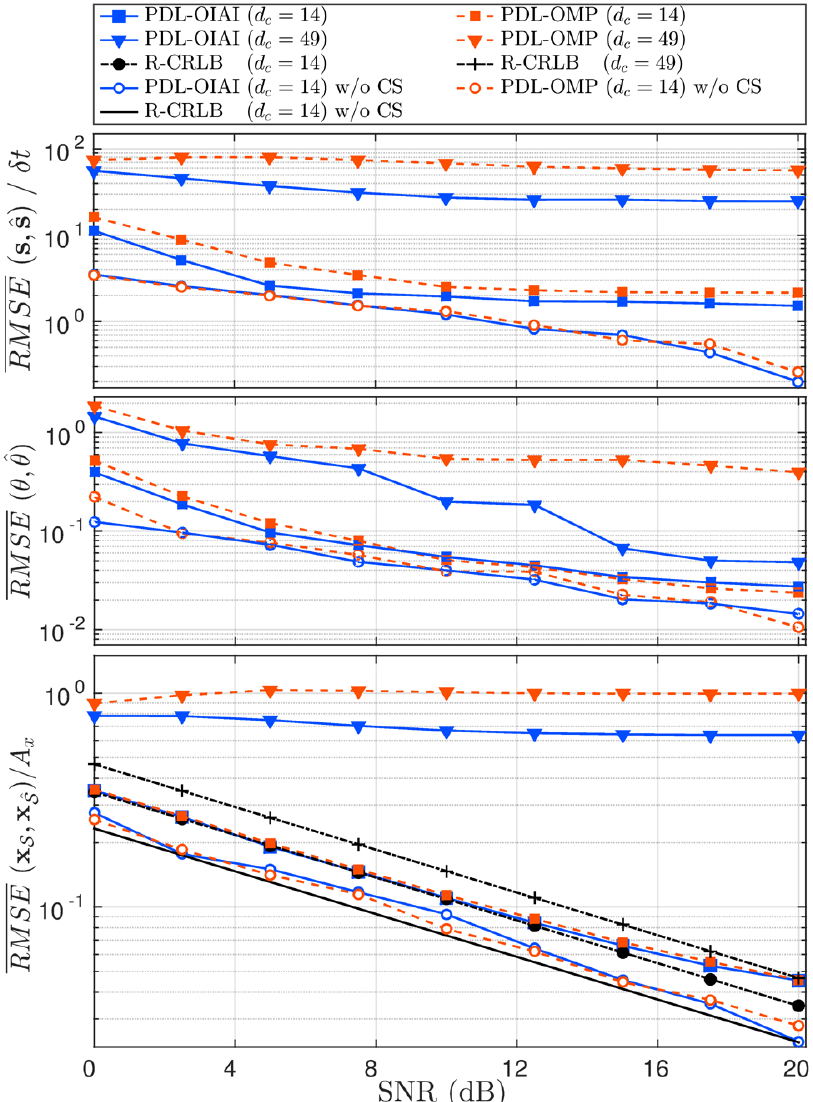}\vspace{1cm}
\caption{Performance of PDL-OIAI and PDL-OMP for dictionaries with different coherence levels, i.e. $d_c = 14$ and $d_c = 49$, using 70\% of the original samples and sampling matrix (a). 
			The results without (w/o) CS for $d_c=14$ serve as a reference.}
  \label{fig:FBG_performance_BW_estimation}  
\end{figure}

\begin{figure}[H]
  \centering\vspace{2cm}
 \includegraphics[width = 0.80\columnwidth]{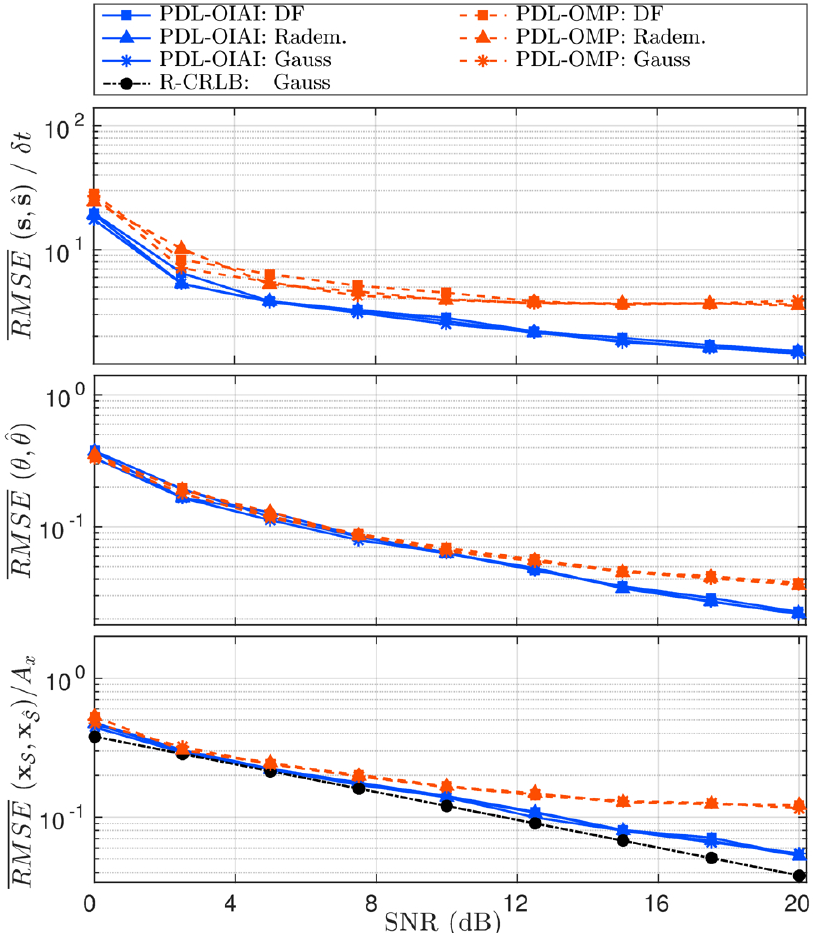}\vspace{1cm}
  \caption{Performance of PDL-OMP and PDL-OIAI using only $50\%$ of the original samples and sampling matrices (a)-(c).}
  \label{fig:performance_different_B_meas_50}
\end{figure}

\begin{figure}[H]
  \centering\vspace{2cm}
\includegraphics[width = 0.8\columnwidth]{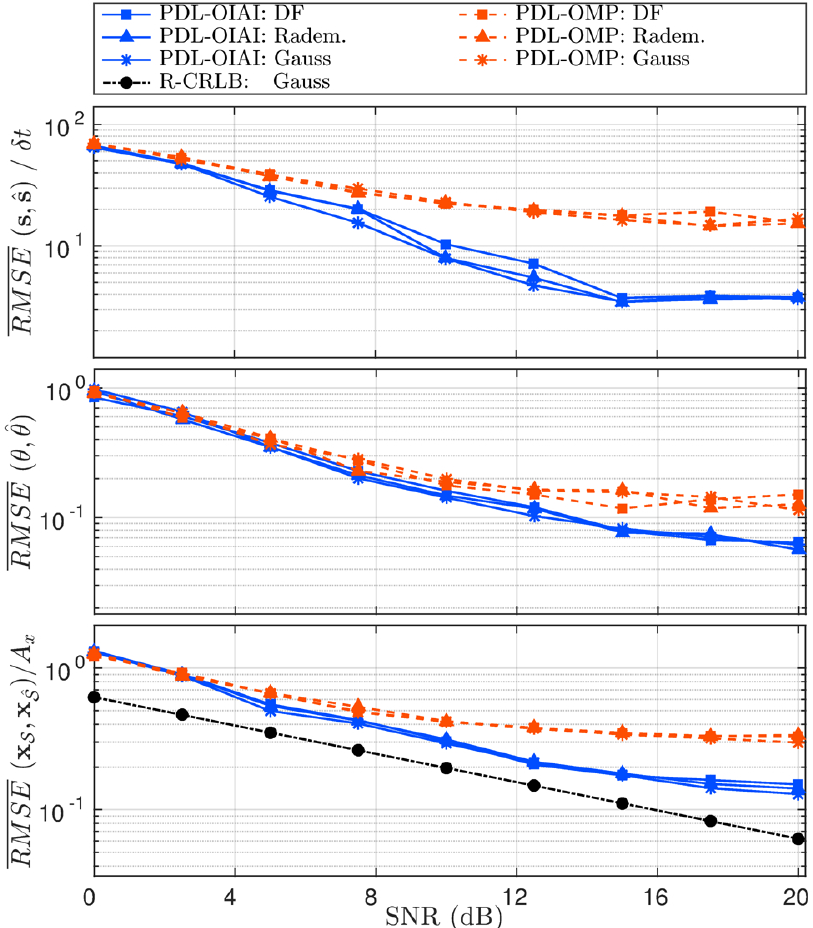}\vspace{1cm}
  \caption{Performance of PDL-OMP and PDL-OIAI using only $20\%$ of the original samples and sampling matrices (a)-(c).}
  \label{fig:performance_different_B_meas_20}
\end{figure}

Next, we investigate the empirical performance of PDL-OIAI and PDL-OMP in order to support the previous observations. %
$\mathcal{S} = \text{supp}\{\mathbf{x}\}$ is the true sparse support and $\hat{\mathcal{S}}$ denotes its estimate.
The vectors $\mathbf{s}$ and $\hat{\mathbf{s}}$ contain, in increasing order, the support and its estimate, respectively.
Simulated sensor data is created using two different values of the effective receiver bandwidth, $\theta_a$ and $\theta_b$, to create two different scenarios
with narrow and broad reflections, respectively. 
The corresponding correctly parametrized dictionaries have coherence distances $d_c =14$ and $d_c = 49$, respectively. 
Recall that for a large value of $d_c$, the dictionary atoms become more similar and it is more difficult to estimate the underlying sparse representation without ambiguity. 
Generally, higher values of $d_c$ require more efforts in IAI mitigation. Therefore, the number of internal iterations in PDL-OIAI is set to $J = 10$ 
when $d_c = 14$, and to $J=18$ when $d_c = 49$. Usually, $J<20$ is sufficient. 
To determine the average performance, we have carried out 500 Monte Carlo trials for each SNR, for each value of $\theta\in[1.2,5]$, and 
for each sampling matrix (a)-(c). We first calculate RMSE$(\mathbf{s},\hat{\mathbf{s}})$ divided by the temporal grid accuracy, $\delta t$. 
It describes the average number of grid points that the estimated support differs from the true support.
Then, we compute RMSE$(\theta,\hat{\theta})$. %
Finally, we compare the estimation performance for the amplitudes of the non-zero coefficients in $\mathbf{x}$ with the 
constrained CRLB in (\ref{eq:constr_CRLB}), which is calculated for the correctly parametrized $\mathbf{B}(\theta)$ using sampling matrix (a). 
Since the constrained CRLB depends on the actual realization of the random matrix $\boldsymbol{\Phi}$, its value is averaged over 1000 Monte Carlo trials.  
Therefore, we only consider the amplitudes in $\mathbf{x}$ at the true positions in $\mathcal{S}$, and the corresponding amplitudes in 
$\hat{\mathbf{x}}$ at the estimated positions in $\hat{\mathcal{S}}$. Hence, we 
define $\mathbf{x}_\mathcal{S} := [x_{s_1},\dots,x_{s_K}]^T = A_x\mathbf{1} \in \mathbb{R}^K$, where
the vector $s_k\in\mathcal{S}, k=1,\dots,K$. $\mathbf{x}_{\hat{\mathcal{S}}}$ is defined accordingly.\\
In Fig.\ \ref{fig:FBG_performance_BW_estimation}, we show the RMSE for estimating $\mathbf{s}, \theta$, $\mathbf{x}_{\mathcal{S}}$, using $M/L=70\%$ of the original measurements. 
For $d_c=14$, the results without (w/o) CS, i.e. $\mathbf{B}=\mathbf{A}$, are shown as a reference.
The smallest estimation error is achieved without CS, i.e. when all original samples are available.  %
The results for RMSE$(\mathbf{s}, \hat{\mathbf{s}})/\delta t$ demonstrate that 
PDL-OIAI yields more accurate estimates for the reflection delays than PDL-OMP at all SNRs. 
The difference becomes more significant as $d_c$ increases but also PDL-OIAI has limited performance when $d_c$ is too severe. 
The values of RMSE$(\theta,\hat{\theta})$ indicate that PDL-OIAI and PDL-OMP both perform well in estimating $\theta$ when $d_c = 14$. 
For $d_c = 49$, and at low SNRs, the influence of noise dominates the performance of both methods, whereas PDL-OIAI is superior at higher SNRs.
Regarding the sparse coefficients, we compare RMSE$(\mathbf{x}_\mathcal{S}, \hat{\mathbf{x}}_{\hat{\mathcal{S}}})$, 
normalized by the common amplitude $A_x$, to the root constrained CRLB (R-CRLB).
The estimates of both algorithms cling closely to the R-CRLB when $d_c = 14$. 
For $d_c = 49$, however, they both do not yield correct estimates due to dictionary coherence.\\
In Fig.\ \ref{fig:performance_different_B_meas_50} and Fig.\ \ref{fig:performance_different_B_meas_20}, we show the performance of PDL-OIAI/OMP for the different sampling matrices (a)-(c) when $d_c = 14$. %
The ratio between the number of CS samples and the number of the original samples is $M/L = 50\%$ in Fig.\ \ref{fig:performance_different_B_meas_50}, 
and only $20\%$ in Fig.\ \ref{fig:performance_different_B_meas_20}. 
When $M/L=20\%$, the sparse support is estimated with an average error of $\overline{\text{RMSE}}(\mathbf{s}, \hat{\mathbf{s}})/\delta t = 4$ bins at SNR$ = 20$ dB. 
The average ratio between the estimation error and the true value of $\theta$ goes down to $\overline{\text{RMSE}}(\theta,\hat{\theta}) = 6\,\%$. 
Comparing the sparse coefficients to the R-CRLB, a maximum performance loss  
of\, $-20\, \text{lg}(\,\overline{\text{RMSE}}(\mathbf{x}_\mathcal{S}, \hat{\mathbf{x}}_{\hat{\mathcal{S}}}) / \text{R-CRLB}\,) = -9$ dB\, 
is observed. %
Although a DF matrix has a high number of zero-projections (2/3 of all samples need not be acquired), 
similar performance is achieved for all sampling matrices. 
\newpage

\newpage

\subsection{Experimental validation using real data}
\label{sec:exp_results}
\begin{figure}[t!]
   \centering
   \includegraphics[width = 0.8\columnwidth]{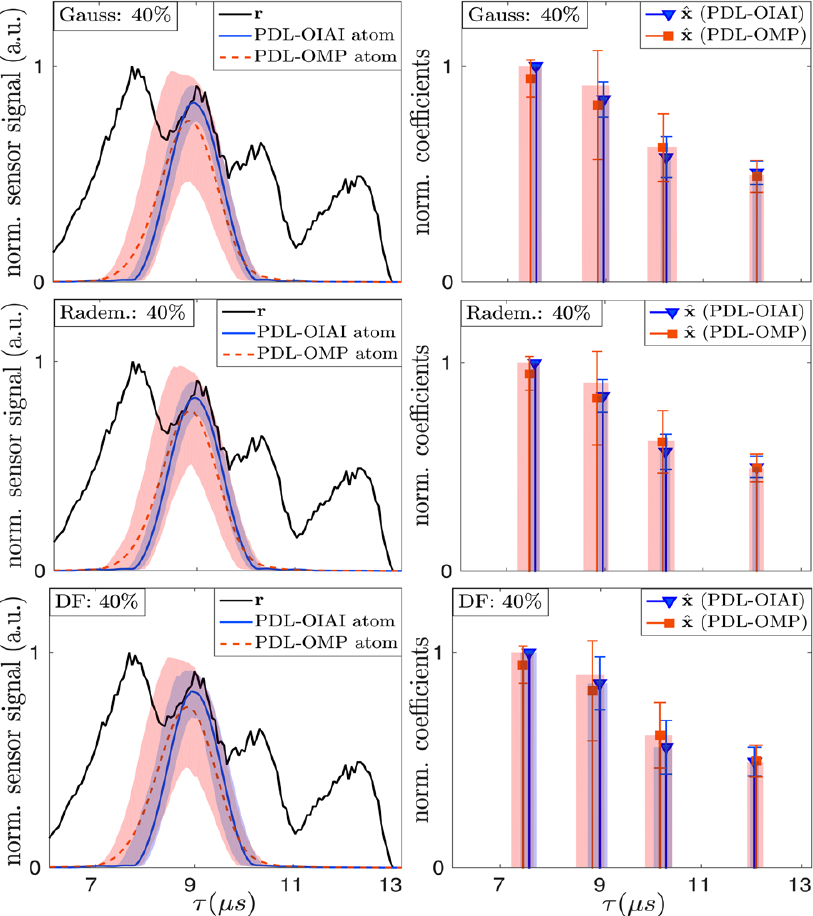}\vspace{0.2cm}
   \caption{Real data example: sensor signal with median of one estimated atom (left), median of the estimated support and amplitudes of $\hat{\mathbf{x}}$ (right).
            For both algorithms, the shaded areas and the error bars correspond to the $85\%$ confidence intervals.}
   \label{fig:realData}
\end{figure}

Next, we apply the truncated CS-DL framework to experimental data taken from  
the reference system in \cite{Yamashita2009,Nakazaki2009}, which was %
acquired at the Yamashita laboratory of photonic communication devices at The University of Tokyo, Japan. 
The basic architecture corresponds to the schematic in Fig.\ \ref{fig:cfsp}, and the total experimental setup is detailed in \cite{Nakazaki2009,Yamashita2009}. 
We tuned the parameters of the model in Section \ref{sec:CS_DL_pratctice} \ref{sec:param_model} to fit the system specifications, 
the measured spectra of laser output and FBG reflections, and the measured laser output. %
We consider $L=134$ samples of the received signal, $\mathbf{r}$, and take CS samples such that $M/L = 40\%$. 
The dictionary has $N = 2L$ atoms with a delay spacing of $\delta t\approx 50$ ns.
The sensing fiber contains $K=4$ FBGs. %
The corresponding reflection delays are approximately at $[7.79, 9.05, 10.27, 12.30]$ $\mu$s 
which is potentially off-grid. In PDL-OIAI, we use $J\leq 8$ iterations for IAI mitigation. To account for the randomness of the sampling matrices, 500 Monte Carlo trials 
were carried out. Ultimately, confidence intervals for the parameters were computed. %
In order to discard outliers that result from an unfortunate conditioning of $\boldsymbol{\Phi}$, we show the results in terms of the median and the $85\%$ confidence 
interval instead of showing mean and standard deviation.
The observed reflections are strongly broadened which, according to the authors in \cite{Nakazaki2009}, probably arises from the limited bandwidth of 
the ADC and/or from an extended laser linewidth. %
In our setting, we relate the remaining uncertainty in the temporal width of the reflections to the effective receiver bandwidth, $\Delta f$. 
Hence, as an auxiliary global parameter, the value of $\Delta f$ that correctly specifies the width of the reflections defines the correct 
dictionary parameter, i.e. $\hat{\theta} = $\text{\small{$\widehat{\Delta f}$}} . %
The total search range for $\theta$ covers dictionaries with 
coherence distances $9\leq \hat{d}_c\leq 89$. The correct value yields dictionaries with $12\leq d_c \leq 30$. %
Besides, there are a few distortions that are not captured by the model in Section \ref{sec:CS_DL_pratctice} \ref{sec:param_model} such as the skew shape of the reflections. 
This may result from a different rise and fall time in the temporal response of the PD. 
Nonetheless, $\mathbf{s}$, $\theta$ and $\mathbf{x}$ can still be estimated, which indicates some robustness to model deviations. %
The rows in \mbox{Fig.\ \ref{fig:realData}} show the performance of PDL-OIAI and PDL-OMP using the sampling matrices (a)-(c), respectively.
The left column depicts the raw sensor signal, $\mathbf{r}$, after photodetection and A/D conversion. For the second reflection, the median of the estimated scaled atoms 
is shown (solid and dashed line). The shaded areas represent the $85\%$ confidence intervals of the atom. %
The right column shows the median of the corresponding sparse coefficients along with the error bars corresponding to the $85\%$ confidence intervals for $\hat{\mathbf{x}}_{\hat{\mathcal{S}}}$.
For the sparse signal, the shaded areas represent the $85\%$ confidence intervals for $\mathbf{s}$. %
Since $d_c$ is not too large and the SNR is relatively high, the \emph{average} performance of both methods is not significantly different. 
Yet, the reflection delays, $\hat{\mathbf{s}}$, identified by PDL-OIAI are more accurate than 
those of PDL-OMP. Also, $\hat{\mathbf{x}}_{\hat{\mathcal{S}}}$, yielded by PDL-OIAI, adheres closer to the amplitudes of the acquired signal.
This is verified by much narrower confidence intervals of the support and amplitude estimates in the case of PDL-OIAI. %
Comparing the data to the estimated atoms, the value of
$\hat{\theta}$, obtained by PDL-OMP, tends to be smaller. The estimated reflections are broader 
than those of the sensor data. For sampling matrices (a)-(c), the medians and $85\%$ confidence regions, $\Delta_{85\%}$, 
of $\hat{\theta}$ are shown in Table \ref{tb:real_data_BW_results}. %
%
%
\begin{table}[b]\vspace{-0.3cm}
 \caption{Medians and 85\%-confidence ranges of the estimated receiver bandwidths $\text{\small$\widehat{\Delta f}$}$ (in MHz) for the experimental data set using CS matrices (a)-(c) with $M/L = 40\%$. }\vspace{0.1cm}
  \label{tb:real_data_BW_results}
\begin{center}
\begin{tabular}{ccccc}   
  \toprule[1.5pt] 
   Algorithm & Measure & Gaussian	& Radem.	& DF  \\[0.2ex] %
%
%
\cmidrule[0.7pt](lr){1-2} \cmidrule[0.7pt](lr){3-5} 
  	 \multirow{2}{2.3cm}{\vspace{-0.0cm}\hspace*{0.0cm} \textnormal{\textbf{PDL-OIAI}}}	&\multirow{2}{1.3cm}{\hspace*{0.10cm}Median\vspace{0.1cm}\\\hspace*{0.25cm}$\Delta_{85\%}$}  & \multirow{2}{1.3cm}{\hspace*{0.15cm}$1.6560$\vspace{0.1cm}\\\hspace*{0.15cm}$0.2565$}	& \multirow{2}{1.2cm}{\hspace*{0.15cm}$1.6555$\vspace{0.1cm}\\\hspace*{0.15cm}$0.2397$}	& \multirow{2}{1.3cm}{\hspace*{0.15cm}$1.6458$\vspace{0.1cm}\\\hspace*{0.15cm}$0.3727$}  \\[3.99ex]%
  	 \multirow{2}{2.3cm}{\vspace{-0.0cm}\hspace*{0.0cm} \textnormal{\textbf{PDL-OMP}}}	& \multirow{2}{1.3cm}{\hspace*{0.1cm}Median\vspace{0.1cm}\\\hspace*{0.25cm}$\Delta_{85\%}$} & \multirow{2}{1.3cm}{\hspace*{0.15cm}$1.6545$\vspace{0.1cm}\\\hspace*{0.15cm}$0.8139$}	& \multirow{2}{1.2cm}{\hspace*{0.15cm}$1.6574$\vspace{0.1cm}\\\hspace*{0.15cm}$0.8115$}	& \multirow{2}{1.3cm}{\hspace*{0.15cm}$1.6371$\vspace{0.1cm}\\\hspace*{0.15cm}$0.7745$} \\[3.7ex] %
  \bottomrule[1.5pt]
\end{tabular}
\end{center}\vspace{-0.5cm}
\end{table}

\section{Discussion}
\label{sec:discussion} %
We show how the CS-DL framework can be truncated to match a particular sensor setup given a model for the received signal.   
In order to apply CS-DL to a particular system, we assume a generic modular architecture and describe the core system in terms of an LTI model. 
This enables the use of CS-DL to a wider range of applications akin to fiber-optic sensing.\\
In our simulations, we show that the reflection delays can be estimated at high resolution using sparse estimation methods.  %
While such methods can be applied without CS, incorporating CS has some additional advantages.  %
First, it reduces the sampling rate and the number of samples to be stored and processed. 
This is desirable since sparse estimation methods are computationally demanding and their complexity increases with the number of samples. 
Second, since $M<L$, the number of rows of the combined sensing dictionary, $\mathbf{B} = \boldsymbol{\Phi}\mathbf{A}$, is determined by the number of measurements, $M$.  
This allows us to control the dictionary coherence by adjusting $M$, where a smaller $M$ yields a higher coherence and increases the difficulty of sparse estimation.
We observe a similar performance for all CS sampling matrices, which 
can be ascribed to the high sparsity level imposed by the proposed dictionary, where only one atom per sub-dictionary can be active by definition.
However, binary sampling matrices are favorable as they are easier to implement in hardware than a Gaussian matrix. %
Uniform sampling can be described by a fully populated sampling matrix, e.g. a Rademacher matrix. %
The sampling rate can be lowered by using several  ADCs in parallel \cite{Strohmer2007}. %
Non-uniform sampling is described by sparse sampling matrices. %
Sparse DF matrices \cite{Achlioptas2003} have a high probability of zero projection (2/3 of all samples need not be acquired), which reduces the average sampling rate by $\approx 66 \%$.  
In some cases, non-uniform sampling can also reduce the hardware costs. 
For example, $P=4$ ADCs are used to sample a uniform grid at rate $f_s$. Then, each ADC can work at the reduced rate $f_s/P$. %
If the sampling grid is sparse and every sequence of 4 samples in the rows of $\boldsymbol{\Phi}$ has either the 
pattern 0100 or 0010 (or a common cyclic shift), then only 2 parallel ADCs at rate $f_s/4$ are sufficient as there are always at least 2 subsequent unoccupied positions.\\ 
It is evident from the previous results that PDL-OIAI is a viable tool for estimating the reflection delays, $\mathcal{S}$, along with the dictionary parameter, $\theta$. 
In addition, the amplitudes of the reflections, i.e. sparse coefficients $\hat{\mathbf{x}}_{\hat{\mathcal{S}}}$, can be estimated with a performance close to the CRLB. 
PDL-OIAI works reliably even though the experimental data exhibits features that are not explicitly modeled (e.g. the skew shape of the reflections).
Different from the methods in \cite{Austin2013,Leigsnering2016,Raja2016}, PDL-OIAI involves dictionary pre-processing and is able to handle severe dictionary coherence by IAI mitigation. 
In our application, the dictionary is highly coherent, such that even optimization-based methods, e.g. $\ell_1$-minimization, are unable to obtain reliable estimates of the underlying sparse representation. 
After dictionary pre-processing, however, a simple OMP algorithm of lower complexity can be used for sparse estimation.
Our comparison between PDL-OMP and PDL-OIAI supports the work in \cite{Yang2010} and further emphasizes the necessity of such pre-processing.
The dictionary we use is model-based and has a composite structure, where a shift-invariant sub-dictionary is assigned to each FBG reflection.
This establishes an immediate relation between the sparse support, $\mathcal{S}$, and the reflection delays to be estimated, which is even preserved after IAI mitigation.
This is a strong advantage to non-parametric dictionary design methods. They can yield a lower coherence level by construction but the atoms have no physical interpretation.\\
To quantify the level of difficulty of sparse estimation, conventional measures, e.g. the \emph{Babel function}, consider only the subset of $K$ dictionary atoms with highest correlation. 
When $K$ is small, they yield similar values for dictionaries with different parametrization. For composite, shift-invariant dictionaries, however, adjacent atoms can be highly correlated but 
only one atom in a group may be active. Therefore, we define the coherence distance, $d_c$, to specify the number of adjacent atoms for which the correlation is still higher than a given threshold. 
For example, $d_c=10$ suggests that two active atoms should have a spacing of at least 10 atoms. \\ %
Next, we mention some limitations of our approach. 
One limitation is that, by subsequently minimizing the residual for individual model parameters, only locally optimal estimates can be obtained. %
If prior knowledge is available by expertise and experience of the operator, a good initial value can be chosen to approach or even attain a global optimum.
Another limitation is that the computational complexity of AM-based estimation quickly increases with the number of parameters to be estimated. 
However, our main goal is to estimate the reflection delays. We show that it is sometimes possible to define an auxiliary parameter that combines the contributions of other parameters. %
In particular, we define the effective receiver bandwidth, $\theta=\Delta f$, which describes the broadening of the reflections.
Finally, the computational cost for IAI mitigation in PDL-OIAI may become another limitation. 
To address this problem, the atoms of $\mathbf{W}$ can be computed in parallel.
Also, our simulations indicate that good performance is achieved when $\mathbf{W}$ is calculated only once in the first OMP iteration.\\
Ultimately, we mention some possible extensions that may improve the estimation performance. 
First, multiple observations can be processed by PDL-OIAI, similar to the method in \cite{Malioutov2005}.
Given that the observations belong to the same stationary perturbation profile of the fiber, this can reduce the sensitivity to noise and also group-sparsity \cite{Eldar2012} can be exploited.
In addition, the grid of $N$ dictionary atoms can be iteratively refined, according to \cite{Malioutov2005}, starting with a small $N$ to reduce the computational load and to improve the precision.%

\section{Conclusion}
\label{sec:conclusion}
We present a generic compressed sampling and dictionary learning framework, called CS-DL. %
CS-DL does not dictate a particular system setup but requires a generative parametric model for the observed signal. 
Imperfect prior knowledge can be incorporated in terms of uncertain global or local parameters. 
We show how CS-DL is truncated to suit a particular system setup for WDM-based quasi-distributed fiber-optic sensing.
To this end, we compile a signal model for the reflected signal. This model is used to generate a sparsity-promoting dictionary for estimating the delays of the reflected sensor signals.
However, the dictionary exhibits strong coherence, which aggravates the sparse estimation performance. 
Since conventional coherence measures cannot be used to distinguish different parametrizations of the proposed dictionary, 
we introduce an auxiliary measure, called the coherence distance.
In order to obtain accurate delay estimates, also the width of the reflections has to be estimated as an unknown dictionary parameter. 
The signal is acquired based on compressed sampling, which can lower the sampling rate and reduce the number of samples to be stored and processed. 
For sparse estimation and dictionary learning, an alternating minimization-based algorithm, called PDL-OIAI, is presented.  
It incorporates an inter-atom-interference mitigation sub-routine to yield a modified sensing dictionary with reduced coherence, thereby alleviating the task of sparse estimation. 
Using simulations and experimental data, we show that PDL-OIAI yields reliable estimates even for a small number of CS samples.
Limitations of our approach are local optima in estimating the model parameters and the computational complexity. 
By parallel processing and by using a fixed sensing dictionary, the computational load can be lowered. 
These issues, and possible extensions of our approach such as multi-measurement processing or grid-refinement, will be addressed in a future investigation.

\newpage

 \bibliographystyle{plain}
\bibliography{./bibliography}


\vspace{0.2cm}
\section{Acknowledgments}
This work was supported by the 'Excellence Initiative' of the German Federal and State Governments 
and the Graduate School of Computational Engineering at Technische Universit{\"a}t Darmstadt.
\noindent 
The authors thank Professor Shinji Yamashita and his group at The University of Tokyo, Japan,
for kindly providing experimental data of the fiber sensor in \cite{Yamashita2009}.

\vspace{0.3cm}

\appendix

\section{CRLB for orthonormal matrices}
\label{app:A}
In order to determine a lower bound for the CRLB with respect to the sensing matrix, $\mathbf{B}$, consider a positive semi-definite 
matrix $\mathbf{H} = \mathbf{B}_{\mathcal{S}}^T \mathbf{B}_{\mathcal{S}}$ with eigenvalues $\lambda_k \geq 0, \ k = 1,\dots,K$. Using Hadamard's inequality, we find 
\begin{equation}
\label{eq:Hadamard_ineq}
  \text{det}(\mathbf{H}) = \prod_{k=1}^{K} \lambda_k \leq \prod_{k=1}^{K} h_{kk} = \prod_{k=1}^{K}\widetilde{\lambda}_k = \text{det}(\mathbf{D}),
\end{equation}
where $h_{kk}$ are the diagonal entries of $\mathbf{H}$ and $\widetilde{\lambda}_k$ are the eigenvalues of some diagonal matrix $\mathbf{D}$. 
Further, for $\mathbf{H}^{-1}$ we find 
\begin{equation}
  \label{eq:lowest_CRLB}
 \text{det}(\mathbf{H}^{-1}) = \prod_{k=1}^{K} \frac{1}{\lambda_k}  \geq  \prod_{k=1}^{K} \frac{1}{\widetilde{\lambda}_k} = \text{det}(\mathbf{D}^{-1})
\end{equation}
Therefore, the lowest CRLB is achieved when $\mathbf{H}$ is a diagonal, i.e. when the columns of $\mathbf{B}$ are orthogonal. Then, $\mathbf{H} = \mathbf{D}$ and 
equality holds in (\ref{eq:Hadamard_ineq}) and (\ref{eq:lowest_CRLB}). Since the columns in $\mathbf{B}$ are normalized, i.e. $\|\mathbf{b}_k\|_2 ^2 = \widetilde{\lambda}_k = 1$, we 
obtain the expression
\begin{equation}
  E\{\|\hat{\mathbf{x}}_{ub} - \mathbf{x}\|_2 ^2\} \geq \sigma_n^2\, \sum_{k=1}^{K}\frac{1}{\widetilde{\lambda}_k} = K\, \sigma_n^2\ .
\end{equation}

\end{document}